\documentstyle[preprint,prc,aps,epsfig]{revtex}
\tighten
\begin{document}
\draft
\title{Antikaon condensation in neutron stars}
\author{Subrata Pal$^1$, Debades Bandyopadhyay$^2$ and Walter Greiner$^1$}
\address{$^1$Institut f\"ur Theoretische Physik, J.W. Goethe-Universit\"at, 
D-60054 Frankfurt am Main, Germany} 
\address{$^2$Saha Institute of Nuclear Physics, 1/AF Bidhannagar, 
Calcutta 700 064, India}

\maketitle

\begin{abstract}
We investigate the condensation of charged $K^-$ meson and neutral $\bar K^0$ 
meson in dense neutron star matter. Calculations are performed in 
relativistic mean field models in which both the baryon-baryon
and (anti)kaon-baryon interactions are mediated by meson exchange. It is 
found that $\bar K^0$ condensation is quite sensitive to the antikaon
optical potential and depends more strongly on the nucleonic equation
of state. For moderate values of antikaon potential and a rather stiff
equation of state, a significant region of maximum mass star will contain
$\bar K^0$ meson. The critical density of $\bar K^0$ condensation is always 
higher than that of $K^-$ condensation. With the appearance of $K^-$ and 
$\bar K^0$ condensates, pairs of $p-K^-$ and $n-\bar K^0$ are produced with 
equal proportion leading to a perfectly symmetric matter of nucleons and 
antikaons in neutron stars. Along with $K^-$ condensate, $\bar K^0$ condensate
makes the equation of state much softer resulting in smaller maximum mass
stars compared to the case without any condensate.
\end{abstract}
\vspace{1cm}

\section{Introduction}

The study of dense matter in laboratories reveals many new and interesting
results. There is a growing interplay between the physics of dense matter and
the physics of compact objects \cite{Li,Bet}.
The composition and structure of a neutron star primarily depends on the 
nature of strong interaction. The equation of state of neutron star matter
encompasses a wide range of densities, from the density of iron nucleus at the 
star's
surface to several times the normal nuclear matter density encountered in the
core. Since the chemical potentials of nucleons and leptons increase
rapidly  with density in the stars interior, several novel phases with
large strangeness fraction may appear. Among these possibilities, hyperon 
formation is quite a robust mechanism. This is possible when the neutron 
chemical potential becomes sufficiently large so that the neutrons at their 
Fermi surfaces can decay into hyperons via weak strangeness non-conserving 
interactions. The threshold densities for hyperon formation strongly 
depend on the nuclear equation of state \cite{Pra97,Gle1}.

In recent years considerable interest has grown in the study of the properties
of kaons, $K$, and antikaons, $\bar K$, in dense nuclear matter and in neutron
star matter. Within a chiral $SU(3)_L\times SU(3)_R$ Lagrangian, it was 
demonstrated by Kaplan and Nelson \cite{Kap} that a negatively charged 
antikaon $K^-$ may undergo
Bose-Einstein condensation in dense baryonic matter formed in heavy ion
collisions. It was further predicted in the chiral perturbation theory that a
$K^-$ condensed phase may form in the core of neutron stars 
\cite{Bro92,Tho,Ell} in consonance with kaon-nucleon scattering data 
\cite{Bro94} and $K^-$ atomic data \cite{Lee}. 
In these chiral Lagrangians, the (anti)kaons are 
directly coupled to nucleons. The strongly attractive $K^-$-nucleon 
interaction increases with density lowering the effective mass $m_K^*$ of 
the antikaons. Consequently, the in-medium energy of $K^-$ meson, 
$\omega_{K^-}$, decreases and $s$-wave $K^-$ condensation sets in when  
$\omega_{K^-}$ equals to the $K^-$ chemical potential $\mu_{K^-}$ 
which, in turn, is equal to the electron chemical potential $\mu_e$ for a cold
catalyzed (neutrino-free) neutron star matter. The typical critical density
for $K^-$ condensation in nucleons-only star matter is about $(3-4)n_0$, where
$n_0$ is the normal nuclear matter density. The exact value, however, depends on
the model used, i.e. on the nucleonic equation of state and on the parameters
employed, especially on the depth of the attractive $K^-$ optical potential. 
The net effect of $K^-$ condensation in neutron star matter is that $K^-$
replaces electrons in maintaining charge neutrality and the energy of
the condensate is lowered because of strongly attractive interactions between
nucleons and $K^-$ mesons. Due to the softening of the equation of state
the masses of the stars are reduced in presence of $K^-$ condensate
\cite{Tho,Mut,Kno,Sch,Gle98}. It was also found that in presence of hyperons,
$K^-$ condensation was delayed to higher density, and may not even
exist in maximum mass stars. Protoneutron (newly born and neutrino rich) stars
with $K^-$ condensate was shown to have maximum masses larger than those of
cold catalyzed (old and neutrino free) stars - a reversal from ordinary
nucleons-only star. These stars may undergo supernovae explosion and then 
could collapse to small mass black holes during deleptonization \cite{Brown}.

The situation for kaons is however quite different.
Theoretical investigations based on Nambu$-$Jona-Lasinio model \cite{Lut},
chiral perturbation theory \cite{Bro94}, and one-boson-exchange model 
\cite{Kno,Sch} yield a repulsive optical potential for 
$K^+$ in nuclear medium. Therefore, $K^+$ condensation is expected to 
be forbidden in neutron stars.

In the traditional meson-exchange picture \cite{Ser}, (anti)kaons and 
baryons interact via the exchange of $\sigma$, $\omega$, and $\rho$ mesons.
In addition to the decrease of $m_K^*$ with density here, the energy 
$\omega_{K^-}$ is 
lowered since $K^-$ meson experiences attractive vector $\omega$-meson 
potential due to G-parity of antikaons. In contrast, the isovector 
$\rho$-meson field
is repulsive for $K^-$ and inhibits $K^-$ condensation delaying it to higher
densities. In neutron stars at densities above normal nuclear matter values, 
most of the protons convert into neutrons by inverse $\beta$-decay since it is 
energetically favorable to posses a small number of electrons. The ratio
of electrons to nucleons which is again equal to the proton fraction due to
charge neutrality is typically $Y_e = Y_p \approx 0.1-0.15$. Therefore,
neutron star matter is highly asymmetric and quite distinct from ordinary
symmetric nuclear matter. This suggests that the $\rho$-meson field should
have a strong repulsive effect on $K^-$ meson. On the other hand, the 
$\rho$-meson
field induces an attractive field for $\bar K^0$ meson which is an 
isodoublet partner of $K^-$ meson. This should lower the effective
energy of $\bar K^0$ meson, $\omega_{\bar K^0}$, compared to that of
$K^-$ meson thereby making $\bar K^0$ meson condensation more
favorable in neutron star matter. The critical density for $s$-wave neutral 
$\bar K^0$ meson condensation is governed by the condition
$\omega_{\bar K^0} = 0$, and should depend sensitively on the equation of
state (EOS). Using variational chain summation method with two-nucleon and 
three-nucleon interactions, Akmal et al. \cite{Akm} recently predicted a
neutral pion condensation phase transition in neutron star matter
at a density of $\sim 0.2$ fm$^{-3}$. 
So far no calculation of neutral $\bar K^0$ condensation and its impact
on the gross properties of neutron stars has been performed. In this paper, we
investigate the effect of antikaon condensation with more emphasize on 
$\bar K^0$ condensate on the constitution and structure of neutron star matter 
in the standard meson exchange model. For this purpose, we employ the
usual relativistic mean field Lagrangian \cite{Ser} for baryons interacting
via meson exchanges and includes also the self-interaction of the mesons.
The antikaons are treated in the same footing as the baryons which interact
by the exchange of the same mesons. For the $K^-$ meson, the Lagrangian
density of Ref. \cite{Gle98} is used, and it is extended to include the
$\bar K^0$ meson. We shall demonstrate within this model that, apart
from the $K^-$ meson (as already studied in Ref. \cite{Kno,Sch,Gle98}),
the $\bar K^0$ condensate may also exist inside a neutron star and has a
significant influence on the star properties. We shall also show that
the threshold densities for the antikaon condensates are quite sensitive
to the poorly known high density regime of the equation of state.

The paper is organized as follows. In section II we describe the 
relativistic mean field (RMF) model of strong interactions. The relevant 
equations for neutron star matter with antikaon condensate are summarized 
in this model. In section III the parameters of the model are discussed and 
effects of antikaon condensates in neutron star matter are presented. 
Section IV is devoted to summary and conclusions.

\section{The Formalism}

The starting point in the present approach is a relativistic field theoretical
model of baryons and (anti)kaons interacting by the exchange of scalar 
$\sigma$, isoscalar vector $\omega$, and vector isovector $\rho$ mesons. The
total hadronic Lagrangian can be written as the sum of the baryonic, kaonic and 
leptonic parts, i.e. ${\cal L} = {\cal L}_B + {\cal L}_K + {\cal L}_l$.
Considering all the charge states of the baryon octet 
$B\equiv \{n,p,\Lambda,\Sigma^+,\Sigma^-,\Sigma^0,\Xi^-,\Xi^0 \}$, the baryonic
Lagrangian is given by
\begin{eqnarray}
{\cal L}_B &=& \sum_B \bar\psi_{B}\left(i\gamma_\mu{\partial^\mu} - m_B
+ g_{\sigma B} \sigma - g_{\omega B} \gamma_\mu \omega^\mu 
- \frac{1}{2} g_{\rho B} 
\gamma_\mu{\mbox{\boldmath $\tau$}}_B \cdot 
{\mbox{\boldmath $\rho$}}^\mu \right)\psi_B\nonumber\\
&& + \frac{1}{2}\left( \partial_\mu \sigma\partial^\mu \sigma
- m_\sigma^2 \sigma^2\right) - U(\sigma) \nonumber\\
&& -\frac{1}{4} \omega_{\mu\nu}\omega^{\mu\nu}
+\frac{1}{2}m_\omega^2 \omega_\mu \omega^\mu
- \frac{1}{4}{\mbox {\boldmath $\rho$}}_{\mu\nu} \cdot
{\mbox {\boldmath $\rho$}}^{\mu\nu}
+ \frac{1}{2}m_\rho^2 {\mbox {\boldmath $\rho$}}_\mu \cdot
{\mbox {\boldmath $\rho$}}^\mu ~.
\end{eqnarray}
Here $\psi_B$ denotes the Dirac spinor for baryon B with vacuum mass $m_B$
and isospin operator ${\mbox {\boldmath $\tau$}}_B$. The scalar 
self-interaction term 
\begin{equation}
U(\sigma) = \frac{1}{3} g_2 \sigma^3 + \frac{1}{4} g_3 \sigma^4 ~,
\end{equation}
is included \cite{Bog} to achieve a realistic compression modulus at 
normal nuclear matter density.

The Lagrangian density for (anti)kaons in the minimal coupling scheme is given
by \cite{Gle98}
\begin{equation}
{\cal L}_K = D^*_\mu{\bar K} D^\mu K - m_K^{* 2} {\bar K} K ~,
\end{equation}
with a covariant derivative 
$D_\mu = \partial_\mu + ig_{\omega K}{\omega_\mu} + i g_{\rho K} 
{\mbox{\boldmath $\tau$}}_K \cdot {\mbox{\boldmath $\rho$}}_\mu$. 
The isospin doublet for the kaons
is denoted by $K\equiv (K^+, K^0)$ and that for the antikaons is  
$\bar K\equiv (K^-, \bar K^0)$. The effective mass of (anti)kaons in this
minimal coupling scheme is given by
\begin{equation}
m_K^* = m_K - g_{\sigma K} \sigma ~,
\end{equation}
where $m_K$ is bare kaon mass. In the mean field approximation (MFA)
adopted here, the meson fields are replaced by their mean values and the
baryon currents by those generated in the presence of the mean meson fields.
For a uniform and static matter within MFA, only the time-like components
of the vector fields, and the isospin 3-component of $\rho$-meson 
field have non-vanishing values. The mean meson fields are denoted by
$\sigma$, $\omega_0$, and $\rho_{03}$.

For $s$-wave (${\bf k}=0$) condensation of antikaons $\bar K$, the 
dispersion relation representing the in-medium energies of 
$\bar K\equiv (K^-, \bar K^0)$ is given by
\begin{equation}
\omega_{K^-,\: \bar K^0} = m_K^* - g_{\omega K} \omega_0 
\mp \frac{1}{2} g_{\rho K} \rho_{03} ~,
\end{equation}
where the isospin projection $I_{3\bar K} =\mp 1/2$ for the mesons 
$K^-$ ($-$ sign) and $\bar K^0$ (+ sign) are explicitly written in the
expression. Since the $\sigma$ and $\omega$ fields generally increase with 
density  and both being attractive for antikaons, the effective energies of 
$\bar K$ are lowered in nuclear medium. Moreover, in nucleons-only matter
$\rho_{03} \equiv n_p - n_n$ ($n_p$ and $n_n$ are the proton and 
neutron densities) is negative, thus the $\rho$-meson field inhibits 
$K^-$ condensation whereas it
favors $\bar K^0$ condensation. Employing G-parity which simply transforms
the sign of vector potential, the energies of kaons
$K\equiv (K^+, K^0)$ are expressed as
\begin{equation}
\omega_{K^+,\: K^0} = m_K^* + g_{\omega K} \omega_0 
\pm \frac{1}{2} g_{\rho K} \rho_{03} ~.
\end{equation}
The $\omega$-meson field being repulsive for the kaons and dominates
over the attractive $\sigma$-meson field at high densities is suggestive
of the fact that kaon condensation should be highly restricted.

The meson field equations in presence of baryons and antikaon condensates 
can be derived from Eqs. (1)-(3) as
\begin{eqnarray}
m_\sigma^2\sigma &=& -\frac{\partial U}{\partial\sigma}
+ \sum_B g_{\sigma B} n_B^S 
+ g_{\sigma K} \sum_{\bar K} n_{\bar K} ~,\\ 
m_\omega^2\omega_0 &=& \sum_B g_{\omega B} n_B
- g_{\omega K} \sum_{\bar K} n_{\bar K} ~,\\ 
m_\rho^2\rho_{03} &=& \sum_B g_{\rho B} I_{3B} n_B 
+ g_{\rho K} \sum_{\bar K} I_{3\bar K} n_{\bar K} ~.
\end{eqnarray}
Here the scalar and number density of baryon $B$ are respectively
\begin{eqnarray}
n_B^S &=& \frac{2J_B+1}{2\pi^2} \int_0^{k_{F_B}} 
\frac{m_B^*}{(k^2 + m_B^{* 2})^{1/2}} k^2 \ dk ~,\\
n_B &=& (2J_B+1)\frac{k^3_{F_B}}{6\pi^2} ~, 
\end{eqnarray}
with effective baryonic mass $m_B^*=m_B - g_{\sigma B}\sigma$,
Fermi momentum $k_{F_B}$, spin $J_B$, and isospin projection
$I_{3B}$. Note that for $s$-wave ${\bar K}$ condensation, the scalar and
vector densities of antikaons are same, and in the MFA is given 
by \cite{Gle98}
\begin{equation}
n_{K^-,\: \bar K^0} = 2\left( \omega_{K^-, \bar K^0} + g_{\omega K} \omega_0 
\pm \frac{1}{2} g_{\rho K} \rho_{03} \right) {\bar K} K  
= 2m^*_K {\bar K} K  ~.
\end{equation}
The total energy density $\varepsilon = \varepsilon_B + \varepsilon_{\bar K}$
has contributions from baryons, leptons, and antikaons. The baryonic 
plus leptonic energy density is
\begin{eqnarray}
\varepsilon_B &=& \frac{1}{2}m_\sigma^2 \sigma^2 + \frac{1}{3} g_2 \sigma^3 
+ \frac{1}{4} g_3 \sigma^4  + \frac{1}{2} m_\omega^2 \omega_0^2 
+ \frac{1}{2} m_\rho^2 \rho_{03}^2  \nonumber \\
&& + \sum_B \frac{2J_B+1}{2\pi^2} 
\int_0^{k_{F_B}} (k^2+m^{* 2}_B)^{1/2} k^2 \ dk
+ \sum_l \frac{1}{\pi^2} \int_0^{k_l} (k^2+m^2_l)^{1/2} k^2 \ dk ~.
\end{eqnarray}
The last term corresponds to the energy of leptons as required in a
neutron star matter. The energy density for antikaons is
\begin{equation}
\varepsilon_{\bar K} = m^*_K \left( n_{K^-} + n_{\bar K^0} \right) .
\end{equation}
Since antikaons form $s$-wave Bose condensates, they do not directly
contribute to the pressure so that the pressure is due to baryons and 
leptons only
\begin{eqnarray}
P &=& - \frac{1}{2}m_\sigma^2 \sigma^2 - \frac{1}{3} g_2 \sigma^3 
- \frac{1}{4} g_3 \sigma^4  + \frac{1}{2} m_\omega^2 \omega_0^2 
+ \frac{1}{2} m_\rho^2 \rho_{03}^2 \nonumber \\
&& + \frac{1}{3}\sum_B \frac{2J_B+1}{2\pi^2} 
\int_0^{k_{F_B}} \frac{k^4 \ dk}{(k^2+m^{* 2}_B)^{1/2}}
+ \frac{1}{3} \sum_l \frac{1}{\pi^2} 
\int_0^{k_l} \frac{k^4 \ dk}{(k^2+m^2_l)^{1/2}} ~.
\end{eqnarray}
The pressure due to antikaons is contained entirely in the meson
fields via their field equations (7)-(9).

At the interior of neutron stars, nucleons and electrons undergo
usual $\beta$-decay processes $n \to p + e^- + \bar \nu_e$ and
$p + e^- \to n + \nu_e$. When the electron chemical potential becomes 
equal to the muon mass, electrons are converted to muons by 
$e^- \to \mu^- + \bar \nu_\mu + \nu_e$. In the present study of cold neutron
star we may assume that the neutrinos have left the system freely. Therefore
the chemical potentials of nucleons and leptons are governed by the equilibrium
conditions
\begin{equation}
\mu_n - \mu_p = \mu_e = \mu_\mu ~,
\end{equation}
where $\mu_n$, $\mu_p$, $\mu_e$, and $\mu_\mu$ are the chemical potentials
of neutrons, protons, electrons, and muons with 
$\mu_{n,p} = (k^2_{F_{n,p}} + m_N^{* 2} )^{1/2} + g_{\omega N} \omega_0
+ I_{3N} g_{\rho N} \rho_{03}$. With the onset of $\bar K$ condensation,
the strangeness changing processes that may occur are 
$N \rightleftharpoons N + \bar K$ and $e^- \rightleftharpoons K^- + \nu_e$,
where $N\equiv (n,p)$ and $\bar K \equiv (K^-, \bar K^0)$ denote the 
isospin doublets for nucleons and antikaons, respectively. The
requirement of chemical equilibrium yields
\begin{eqnarray}
\mu_n - \mu_p &=& \mu_{K^-} = \mu_e ~, \\
\mu_{\bar K^0} &=& 0 ~,
\end{eqnarray}
where $\mu_{K^-}$ and $\mu_{\bar K^0}$ are respectively the chemical
potentials of $K^-$ and $\bar K^0$. The above conditions dictate the
onset of antikaon condensations. When the effective energy of $K^-$ meson,
$\omega_{K^-}$, equals to its chemical potential, $\mu_{K^-}$, which in
turn is equal to the electrochemical potential $\mu_e$, a $K^-$ condensate 
is formed. While $\bar K^0$ condensation follows when its in-medium 
energy satisfies the condition $\omega_{\bar K^0} = \mu_{\bar K^0} = 0$. 
When other baryons in the form
of hyperons are present in the neutron star matter, the standard 
$\beta$-decay processes for nucleons generalize to the form
$B_1 \to B_2 + l + \bar \nu_l$ and $B_2 + l \to B_1 + \nu_l$, where $B_1$ 
and $B_2$ are baryons and $l$ is a lepton. All the equilibrium conditions
involving the baryon octet may then be summarized by a single generic
equation
\begin{equation}
\mu_B = \mu_n - q_B\mu_e ~,
\end{equation}
where $\mu_B$ and $q_B$ are, respectively, the chemical potential and 
electric charge of baryon species $B$. The above relations indicate that,
in chemical equilibrium, two independent chemical potentials, $\mu_n$ and
$\mu_e$, exist which correspond to baryon number and electric charge
conservation. For neutron star matter we need to include also the 
charge neutrality condition, which in presence of antikaon condensate is
expressed as
\begin{equation}
\sum_B q_B n_B - n_{K^-} - n_e  - n_\mu = 0 ~.
\end{equation}

In the present calculation we consider antikaon condensation
as a second order phase transition. In principle in neutron star matter
with two conserved charges, baryon and electric charge, the condensation 
can also be treated as a first order phase transition \cite{Gle98}. In
this situation, the star would have at low density a normal phase
of baryons and leptons followed by a mixed phase of $K^-$ condensate
and baryons, and possibly a pure $K^-$ condensed phase at a higher
density. However in the present situation this would be more complicated 
as $K^-$ and $\bar K^0$ have to be treated as two separate phases, and 
especially when the threshold density of one condensate may 
be reached in the mixed phase of nucleons and the other condensate.
Treatment of such a triple phase would be numerically very involved
and is postponed to a future publication.

\section{Results and discussions}

In the effective field theoretic approach adopted here, three distinct sets
of coupling constants for nucleons, kaons, and hyperons associated
with the exchange of $\sigma$, $\omega$, and $\rho$ mesons are required.
The nucleon-meson coupling constants generated by reproducing the
nuclear matter saturation properties are taken from Glendenning and 
Moszkowski of Ref. \cite{Gle91}. This set is referred to as GM1 and
listed in Table I.

Let us now determine the kaon-meson coupling constants. According to
the quark and isospin counting rule, the vector coupling constants are
given by
\begin{equation}
g_{\omega K} = \frac{1}{3} g_{\omega N} ~~~~~ {\rm and} ~~~~~
g_{\rho K} = g_{\rho N} ~.
\end{equation}
The scalar coupling constant is obtained from the real part of the 
$K^-$ optical potential at normal nuclear matter density 
\begin{equation}
U_{\bar K} \left(n_0\right) = - g_{\sigma K}\sigma - g_{\omega K}\omega_0 ~.
\end{equation}
The negative sign in the vector meson potential is due to G-parity.
The critical density of $\bar K$ condensation should therefore strongly
depend on the $K^-$ optical potential. Recent fits \cite{Li,Fri94,Fri99} 
to the $K^-$ atomic data indicate a strong attractive potential in dense
matter. More recently, in high energy heavy ion collisions enhanced
subthreshold $K^-$ production with a steep spectral slope was found 
\cite{Lau}. In a simplified model this may be interpreted as a strong
$K^-$ attraction in nuclear matter.
On the other hand, available $K^-N$ scattering length suggests
a repulsive interaction. The reason for this apparent ambiguity is the 
presence of $\Lambda(1405)$-resonance which is considered to be an unstable
${\bar K} N$ bound state just below the $K^- p$ threshold that makes the 
interpretation of the data more complicated. Recent analysis \cite{Fri99} of 
$K^-$ atomic 
data using a hybrid model which combines the relativistic mean field approach
in the nuclear interior and a phenomenological density dependent potential
at low density showed that the real part of $K^-$ optical potential could 
be as large as $U_{\bar K} = -180 \pm 20$ MeV at normal nuclear matter 
density while being slightly repulsive at low density in accordance 
with the low density theorem. 
Coupled channel formalism which automatically
generates the $\Lambda(1405)$-resonance and successfully describes the low
energy $K^- p$ scattering data yields an attractive potential for the $K^-$ 
meson of $U_{\bar K}(n_0) \approx -100$ MeV \cite{Koc}. In a chirally motivated
coupled channel approach, the $\bar K$ optical potential depth was found 
to be $U_{\bar K}(n_0) = -120$ MeV \cite{Waa}. The reason for such a spread
in the predicted values lies in the diversity of treating the 
$\Lambda(1405)$-resonance. We have therefore determined the $K-\sigma$ 
coupling constant $g_{\sigma K}$ for a set of values of 
$U_{\bar K}(n_0)$ starting from $-100$ MeV to $-180$ MeV. This is listed
in Table II for the set GM1. Since the $\omega$-meson potential for $\bar K$
in this model is $V_\omega^K(n_0) = g_{\omega K} \omega_0 \approx 72$ MeV,
a rather large sigma-kaon coupling constant of $g_{\sigma K} = 3.674$ is
required to reproduce a depth of $-180$ MeV. Note that only for this large depth
the value of the scalar coupling is similar to the prediction in the
simple quark model i.e., $g_{\sigma K} = g_{\sigma N}/3$.

As an alternative approach the kaon-meson coupling constants may be determined
from the $s$-wave kaon-nucleon ($KN$) scattering length. At the tree
level, the isospin averaged $KN$ scattering length is given by \cite{Coh}
\begin{equation}
{\bar a}_{KN} = \frac{1}{4}a_0^{I=0} + \frac{3}{4}a_0^{I=1}
= \frac{m_K}{4\pi(1+m_K/m_N)} \left( \frac{g_{\sigma K}g_{\sigma N}}{m^2_\sigma}
- 2\frac{g_{\omega K}g_{\omega N}}{m^2_\omega} \right) .
\end{equation}
From the low density theorem, the kaon optical potential depth is
given by
\begin{equation}
U_K=- \frac{2\pi}{m_K} \left( 1+\frac{m_K}{m_N} \right) {\bar a}_{KN}\: n_B ~.
\end{equation}
Using the experimental values \cite{Bar} of $a_0^{I=1} = -0.31$ fm and
$a_0^{I=0} = -0.09$ fm, the scattering length is ${\bar a}_{KN} = -0.255$ fm.
The $K$ optical potential depth is then $U_K \approx +29$ MeV at $n_0=0.153$ 
fm$^{-3}$ and is repulsive. In the GM1 parameter set, for $\bar K$ depths
of $U_{\bar K}(n_0) = -100 (-180)$ MeV, we obtain $KN$ average scattering 
lengths and $K$ optical potential depths of ${\bar a}_{KN} = -0.468 (-0.319)$ 
fm and
$U_K(n_0) \approx +44 (-36)$ MeV. With $U_{\bar K}(n_0) = -180$ MeV, 
the attractive potential obtained for kaon optical potential 
depth (apart from rather 
small vector potential of (anti)kaon $V_\omega^K$) is due to the neglect of 
$\Lambda(1405)$-resonance which is beyond the scope of this paper.

In the present study we shall be mostly concerned with antikaon condensation
in nucleons-only matter. In general, the presence of hyperons delays the onset
of $\bar K$ condensation to much higher densities \cite{Ell,Mut,Kno,Sch}. 
However, for orientation we shall only provide briefly the effects of 
hyperons on antikaon condensation in neutron star matter. The vector coupling 
constants for the hyperons are obtained from $SU(6)$ symmetry
\begin{eqnarray}
\frac{1}{3} g_{\omega N} &=& \frac{1}{2} g_{\omega \Lambda}
= \frac{1}{2} g_{\omega \Sigma} = g_{\omega \Xi} ~, \nonumber \\
g_{\rho N} &=& \frac{1}{2} g_{\rho \Sigma} = g_{\rho \Xi} ~; 
~~~~ g_{\rho \Lambda} = 0 ~.
\end{eqnarray}
The $\sigma$-meson couplings to the hyperons ($Y$) can be obtained from
the well-depth of $Y$ in saturated nuclear matter
\begin{equation}
U_Y^N \left(n_0\right) = - g_{\sigma Y} \sigma + g_{\omega Y}\omega_0 ~.
\end{equation}
Analysis of energy levels in $\Lambda$-hypernuclei suggests \cite{Chr} a 
well-depth of $\Lambda$ in nuclear matter of $U^N_{\Lambda}(n_0)\approx - 30$ MeV.
For the $\Sigma$ potential the situation is unclear, since there is no 
evidence for bound $\Sigma$-hypernuclei. The prediction range from completely
bound $\Sigma$'s \cite{Dov} with $U^N_{\Sigma}(n_0)\approx - 30$ MeV to 
unbound \cite{Fri94} with $U^N_{\Sigma}(n_0)\approx + 30$ MeV.
A few events in emulsion experiments with $K^-$ beams have been attributed 
\cite{Chr} to the formation of $\Xi^-$-hypernuclei from which the depth 
of $\Xi$ in nuclear matter has been extracted to be 
$U^N_\Xi (n_0) \approx - 28$ MeV. Recently, a few $\Xi$-hypernuclei events
have been identified in ($K^-,K^+$) reaction from which a tighter constraint on
$\Xi$ well-depth of $\approx -18$ MeV has been imposed \cite{Fak,Kha}. 
The coupling constants $g_{\sigma Y}$ for 
hyperons have been obtained using these depths.

We now present results for neutron star matter containing nucleons, leptons, and
$\bar K$ condensates for the parameter set GM1 of Tables I and II. In Fig. 1, 
the nucleon scalar and vector potentials, and the electron chemical potential
are displayed as a function of baryon density normalized to the equilibrium
value of $n_0$ for $K^-$ optical potential depth of $U_{\bar K}(n_0)=-120$ MeV.
The solid lines refer to results for nucleons-only matter and the dashed ones
correspond to calculation with $\bar K$ condensate. With $\bar K$ condensation
we find that the scalar potential is enhanced while vector $\omega$- and
$\rho$-meson potentials are decreased. The kinks in the dashed lines
correspond to the onset of $\bar K^0$ condensation when the meson potentials
deviate further from nucleons-only results. The variation in the meson
fields may be attributed to the occurrence of the source terms due to
$\bar K$ condensation in the field equations of motion (Eqs. (7)-(9)).

The populations of neutron star matter with and without $\bar K$ condensation
are shown in Fig. 2. In the top panel, the particle abundances of nucleons-only
matter are shown. Here all the particle fractions increase with baryon density.
The behavior of the proton (and lepton) fractions are determined by nuclear
symmetry energy which in turn is controlled by the $\rho$-meson field in the
RMF models. In the central panel we exhibit the particle
abundances of the star matter with $K^-$ condensation at $U_{\bar K}(n_0)=-120$
MeV, which in fact is the first particle to condense among the two antikaons
$\bar K \equiv (K^-, \bar K^0)$. Once $K^-$ condensate sets in at $3.05n_0$,
it rapidly increases with density replacing the leptons in maintaining
charge neutrality. The proton density which then becomes equal to $K^-$ density
eventually turns out to be larger than the neutron density; the neutron density 
nearly freezes to a constant value. This can be understood from the 
threshold condition of $K^-$ condensation $\omega_{K^-} = \mu_n - \mu_p$ 
(see Eq. (17)). Substituting explicitly these values, we obtain
\begin{equation}
\left( k_{F_p}^2 + m_N^{* 2} \right)^{1/2}
= \left( k_{F_n}^2 + m_N^{* 2} \right)^{1/2}
+ \left( -m_K^* + \frac{1}{3} g_{\omega N}\omega_0
- \frac{1}{2} g_{\rho N}\rho_{03} \right) .
\end{equation}
Since at high densities, the $\omega$-meson potential dominates the 
$\sigma$-meson
potential and $\rho_{03} \equiv n_p - n_n$ is negative, the term
within the bracket is positive. It is therefore evident that 
$k_{F_p} > k_{F_n}$. This has been also observed in Ref. \cite{Gle98}. 
Hence the net effect of $K^-$ 
condensation is to lower  the electron chemical potential $\mu_e$ and the
nuclear symmetry energy (see also Fig. 1). In the bottom panel of Fig. 2, we
show the particle fractions with both $K^-$ and $\bar K^0$ condensates.
Interestingly, with the onset of $\bar K^0$ condensate at $4.83n_0$, the
neutron and proton abundances become identical resulting in an isospin
saturated symmetric nuclear matter. This can be easily understood by 
substituting the threshold condition for $\bar K^0$ condensation, i.e.
$\omega_{\bar K^0} = 0$, from Eq. (5) into Eq. (27). From these relations 
it is clear that $\bar K^0$ condensate formation enforces the 
condition $k_{F_p} = k_{F_n}$. Alternatively, one may interpret from the 
central panel of Fig. 2 that once the neutron and proton densities 
(Fermi momenta) become identical, the isodoublet partner
of $K^-$ meson, i.e. $\bar K^0$ should be populated. The electrochemical
potential $\mu_e = \mu_n - \mu_p$ then merges with the magnitude of the 
$\rho$-meson potential $|V_\rho^N| = |g_{\rho N} \rho_{03}|$ with
$\rho_{03} \equiv n_{\bar K^0} - n_{K^-}$, as is evident from Fig. 1. 
At densities
above $7.32n_0$, the antikaon densities are equal which is a general feature
of isospin driving force towards symmetry. The $\rho$-meson field and thereby
$\mu_e$ vanishes. As a consequence we are left with a system composed of
exactly equal amount of $n-p-{K^-}-\bar K^0$, i.e. a perfectly symmetric
matter of nucleons and antikaons. For this matter, the strangeness fraction
$f_S = |S|/B = (n_{K^-} + n_{\bar K^0})/n_B$ is as large as unity.

In Fig. 3, the effective masses of the (anti)kaons, $m_K^*$, are shown as a 
function of normalized density for various values of $K^-$ optical potential 
depths
from $-100$ MeV to $-180$ MeV in steps of 20 MeV. Note that $\bar K$ is only a
test particle in the field of nucleons and appear physically only after
condensation. The decrease of $m_K^*$ is found to be quite sensitive to
the $K^-$ depth which determines the onset of condensation. At a given
$U_{\bar K}(n_0)$, the (anti)kaon mass decreases more strongly 
with successive condensation of $K^-$ and $\bar K^0$ mesons.
This is a manifestation of increase in the $\sigma$-meson field with
condensation as can be seen in Fig. 1.

Figure 4 shows the $s$-wave condensation energies for $K^-$ meson,
$\omega_{K^-}$ (solid lines), and for $\bar K^0$ meson, $\omega_{\bar K^0}$
(dashed lines), as a function of density of various antikaon optical potential 
depths in the GM1 set. The effective energies follow a similar 
qualitative trend as $m_K^*$. However, for depths
up to $U_{\bar K}(n_0)=-140$ MeV, the effective kaon masses do not vary
significantly (see Fig. 3). Therefore, considering Eq. (5), the large drop
in $\omega_{K^-,\: \bar K^0}$ with density is dominated by $\omega$-meson
that influences $\bar K$ condensation. For higher values of $U_{\bar K}$,
the large decrease in $m_K^*$s due to large couplings 
$g_{\sigma K}$ are primarily
responsible for condensation. For any value of $U_{\bar K}$, the
isovector potential of nucleons shifts the energy of $\bar K^0$ meson
below that of the $K^-$ meson by $\sim 120$ MeV. At a given $U_{\bar K}$,
condensation of $K^-$ meson occurs when $\omega_{K^-}$ intersects the
electron chemical potential $\mu_e$ in absence of a condensate 
(shown by dotted line). For all values of $\bar K$
depth we find that $K^-$ meson is formed before the threshold density 
for $\bar K^0$ condensate, i.e. $\omega_{\bar K^0}=0$, is reached. The critical
densities for $\bar K$ condensates are collected in Table III for different
values of $U_{\bar K}(n_0)$. A careful investigation of Fig. 4 reveals that
beyond the densities for $K^-$ condensates (and before $\bar K^0$ threshold
densities), the rate of decrease of the energies $\omega_{K^-,\: \bar K^0}$
are attenuated. This can be traced back to the density-dependent behavior 
of the meson fields in Fig. 1. Above $K^-$ formation density the decrease in
$\omega$-meson field dominates over the increase in the $\sigma$-meson field.
This leads to smaller rate of decrease of $\omega_{K^-}$ and 
$\omega_{\bar K^0}$. Moreover, the decrease in the $\rho$-meson field 
due to $K^-$ condensation causes even a smaller decrease of $\bar K^0$ 
energy with density. Thus the $K^-$ meson condensate with the help of
the vector mesons  conspires to delay the appearance of its isospin partner
$\bar K^0$ meson to a higher density. At densities above $\bar K^0$ condensate,
the $K^-$ energy again starts to decrease due to dramatic fall of $\rho$-meson
potential.

The equation of state (EOS), pressure $P$ versus the energy density 
$\varepsilon$ is displayed in Fig. 5 for the GM1 parameter set. The solid line
represents calculation for nucleons-only star matter while the dashed lines 
refer to those with $\bar K$ formation for different values of 
$U_{\bar K}(n_0)$. The 
strong attraction imparted by antikaon condensation makes the EOS softer. The
kinks at higher densities in the dashed lines correspond to the appearance
of $\bar K^0$ condensation which further soften the EOS. Also the softness is
quite sensitive to the choice of the $\bar K$ optical potential depth. A large 
$U_{\bar K}(n_0)$ corresponds to larger attraction and thereby an
enhanced softening of the EOS.
 
We have used the results of Baym, Pethick and Sutherland \cite{Bay} to describe
the crust consisting of leptons and nuclei at the low-density ($n_B<0.001$
fm$^{-3}$) EOS. For the mid-density regime ($0.001 < n_B < 0.08$ fm$^{-3}$)
the results of Negele and Vautherin \cite{Neg} are employed.
Above this density, the EOS for the relativistic models have been adopted.
It is worth mentioning here that though we have treated $\bar K$ condensation
as a second order phase transition, for higher $U_{\bar K}(n_0)$ values
the incompressibility $K = 9 dP/dn_B$ at the threshold is found to be 
negative (see Fig. 5). This represents a first order phase transition
where we have employed the Maxwell construction to maintain a positive 
compressibility.

The static neutron star sequences representing the stellar masses $M/M_\odot$ and
the corresponding central energy density $\varepsilon_c$ are shown in Fig. 6
for nucleons-only matter (solid line) and matter with further inclusion of
$\bar K$ condensate (dashed line) for several values of $U_{\bar K}(n_0)$.
With the occurrence of $\bar K$ condensation, the maximum masses of neutron
stars are reduced due to softening of the EOS. The maximum masses $M_{\rm max}$
and the central densities $u_{\rm cent} = n_{\rm cent}/n_0$ of the stars 
are listed in Table III.
Interestingly, it is observed that for $U_{\bar K}(n_0) \geq -140$ MeV, the
threshold densities for neutral $\bar K^0$ condensate occur before the
central densities for the maximum mass stars. This implies that a significant
region of these maximum mass stars will contain $\bar K^0$ condensate along 
with $K^-$ condensate. Note that for large values of $U_{\bar K}(n_0)$, the
requirement of Maxwell construction causes a negligible variation of the 
masses (and radii) of the stars with central density.

Apart from the $\bar K$ optical potential depth, the critical densities for 
antikaon
condensation depend sensitively on the nuclear equation of state at high
density. The GM1 set is obtained from a Walecka-type Lagrangian with 
a self-interaction term for the scalar meson. Because of the linear 
density-dependence of the vector $\omega$-meson field, it was found that the
vector potential overestimates the relativistic Bruckner-Hartree-Fock (RBHF)
results at high densities. Bodmer \cite{Bod} first proposed an additional
nonlinear $\omega$-meson term in the RMF model of the form
\begin{equation}
{\cal L}_{\omega^4} = \frac{1}{4} g_4 \left(\omega_\mu \omega^\mu \right)^2 ~.
\end{equation}
Later, it was found by Sugahara and Toki \cite{Sug} that this modification
led to a reasonable agreement with the RBHF results. The model
parameters obtained \cite{Sug} by fitting experimental data for binding
energies and charge radii of heavy nuclei is referred to as TM1. The
parameters of this set are given in Table I. 
The $\sigma-K$ coupling constants $g_{\sigma K}$
for the TM1 set listed in Table II are found to be much smaller than GM1. This 
stems from rather large $\omega-K$ potential of $V_\omega^K(n_0) = 91$ MeV for
the TM1 set compared to 72 MeV for GM1. In the TM1 set, for $\bar K$ depths
of $U_{\bar K}(n_0) = -100 (-180)$ MeV, we obtain $KN$ average scattering 
lengths and kaon optical potential depths of ${\bar a}_{KN} = -0.829 (-0.373)$ 
fm and
$U_K(n_0) \approx +82 (+3)$ MeV. In this set even for $U_{\bar K}(n_0)=-180$
MeV, the kaon optical potential depth is repulsive due to rather large 
$V_\omega^K(n_0)$ in contrast to GM1 set.

The equation of state for nucleons-only star matter with and without antikaon
condensation is shown in Fig. 7 for the TM1 set. In contrast to the GM1
set, the EOS for $np$-matter at high densities is much softer here. This is
due to $n_B^{1/3}$ variation of the vector potential compared to simple $n_B$
dependence in the GM1 set. Due to a soft EOS, the meson fields in the TM1 set 
vary rather slowly with density leading to delayed appearance of $\bar K$ 
condensate (see Table III). The stellar sequence for this set is shown in
Fig. 8. It is found that the masses of the stars with only nucleons and
leptons are smaller than those in GM1. On the contrary, with $\bar K$ 
condensate the $M_{\rm max}$ of stars
for $U_{\bar K}(n_0)\geq -120$ MeV are in fact larger in the TM1 model
compared to the GM1 set. The delayed formation of $\bar K$ in this model
suppresses the softening effect of antikaons. In particular, only for 
$\bar K$ depth as large as $-160$ MeV, a $\bar K^0$ condensate can be formed
in the limiting mass star.

The above study suggests that a stiffer EOS is more efficient for antikaon
condensation in neutron star matter. The small values of $n_0=0.145$ fm$^{-3}$
and $m_N^*/m_N = 0.634$ with large $a_{\rm asy} = 36.9$ MeV in the TM1 set
should have, in principle, provided a stiffer EOS \cite{Gle1}. However, 
the nonlinear $\omega$-meson term in this model entails a 
substantial softening of the EOS at high
densities. To investigate the effects of a rather stiff EOS on $\bar K$
condensation with acceptable nuclear matter properties, we have generated
another set of parameters using the Lagrangian of Eq. (1) (i.e. excluding
the nonlinear $\omega$-meson term of Eq. (28)) and reproducing the saturation 
properties of the TM1 set. We call this set as GMT and it is presented in
Table I. It is to be noted that the coefficients $g_3$ in the sets GM1 and
GMT are negative. This is associated with the well-known problem that the
scalar potential is unbounded from below. We have however found that in these
effective field theoretical models the equations of state for the two sets 
are continuous (as are exhibited) over a wide density range relevant 
to the neutron star interior. The $g_{\sigma K}$ coupling constants 
presented in Table II for the GMT set are nearly
the same as TM1. This arises from similar values of $\omega-K$ potential
$V_\omega^K$ at saturation density although they should differ considerably
at higher densities. For the same reason, the values of $KN$ scattering
lengths and $K$ optical potential depths at $n_0$ are almost identical 
in the TM1 and GMT sets.

In Fig. 9, we display the variation of effective kaon mass $m_K^*$ (top panel)
and $\bar K$ condensate energies (bottom panel) for various values of
$U_{\bar K}(n_0)$ for the GMT set. This should be contrasted with 
Figs. 3 and 4 for the 
GM1 set. It is found that $m_K^*$ has a smaller decrement with density even
for $U_{\bar K}(n_0)=-180$ MeV. Therefore, the variation of $\omega$-meson
field is chiefly responsible for the large drop in $\omega_{K^-,\: \bar K^0}$
in the GMT set. The EOS and the stellar sequences are shown in Figs. 10 and 11
with and without $\bar K$ condensation. The results for the GMT set are
summarized in Table IV. The $M_{\max}$ for nucleons-only stars is largest 
in the GMT set. Since the EOS is very stiff, the chemical potentials for 
nucleons increase rapidly with density which shift the threshold densities for 
$\bar K$ condensation to much smaller values. We find that $\bar K^0$
condensation occurs well inside the maximum mass stars for all values of 
$U_{\bar K}(n_0)$, except for $-100$ MeV. To delineate the effects of
$\bar K^0$ condensation from $K^-$, we also give in Table IV the maximum
masses and corresponding central densities for stars allowing only $K^-$
condensation. It is found that the softening due to $\bar K^0$ condensate
could reduce the maximum mass from that of the $K^-$ condensate further
by $\approx 17\%$ at $U_{\bar K}(n_0)=-180$ MeV. The present investigation
points to the fact that a stiff EOS (as for the GMT set) favors strongly
the formation of $\bar K^0$ condensate in neutron stars. In other words,
from G-parity one can infer that a soft EOS would enhance kaon production.
Indeed, more kaon production was found \cite{Aic,Har,Li95} in heavy ion 
collisions for a soft EOS rather than a stiff one.

We have discussed in Fig. 3 that $K^-$ condensation which occurs at earlier
density than $\bar K^0$ reduces the magnitude of the vector potentials. 
Consequently, the effective energy $\omega_{\bar K^0}$ of $\bar K^0$ is
enhanced shifting its threshold to a higher density. This effect is
illustrated in Table V where the stellar properties of stars having only
$\bar K^0$ condensate is considered. It is evident that $\bar K^0$ formation 
takes place much earlier in the absence of $K^-$ condensate (see Tables III 
and IV). Since $u_{\rm cr} (\bar K^0) < u_{\rm cent}$, sizeable amount
of $\bar K^0$ condensate would be present in stars for all $\bar K$ optical
potential depths in the GM1 and GMT sets. However, $u_{\rm cr}(\bar K^0)$ 
being higher than $u_{\rm cr}(K^-)$, and moreover, $\bar K^0$ cannot replace 
leptons in maintaining charge neutrality, the EOS is stiffer and the maximum 
masses of the stars are larger here than those with $K^-$ condensate only.

In Fig. 12, the mass-radius relationship is shown for nucleons-only stars
and for stars with different values of $U_{\bar K}(n_0)$ for the GM1, TM1,
and GMT sets. It is found that in all models the limiting mass stars
without $\bar K$ condensate have the largest maximum masses with smallest
radii. With increasing $\bar K$ optical potential depth although the maximum 
masses
decrease, the corresponding radii are however found to be similar for most
of the stars in the TM1 and GMT sets. In contrast, stars with smaller masses
have larger radii in the GM1 set.

We now briefly consider the case when hyperons are allowed in addition to
nucleons. At higher densities when the Fermi energy of nucleons exceeds
the effective mass of a hyperon minus its associated interaction, the 
conversion of nucleons to hyperon is energetically favorable. The
hyperon-meson coupling constants are obtained as discussed before. Unless
otherwise mentioned we use an attractive $\Sigma$ well-depth of $-30$ MeV.
With this choice the $\Lambda$ and
$\Sigma^-$ are the first two strange particles that appear at roughly the same
density of $2n_0$ in all the sets. More massive and positively charged 
particles than these appear at higher densities. Since the conversion of 
nucleons to hyperons relieves the Fermi pressure of nucleons, the equation
of state is softened. The EOSs for star matter with only nucleons and hyperons
are depicted by dash-dotted lines in Figs. 5 and 7 for the GM1 and TM1 sets.
The maximum masses and the corresponding central densities are given in 
Table III. It is evident from the figures that the hyperons emerge at densities 
much earlier than the threshold density for $\bar K$ condensate even for
$U_{\bar K}(n_0)=-180$ MeV. Since we have seen that a soft EOS prevents
$\bar K$ condensation, the hyperons therefore postpone $\bar K$ formation
to higher densities. This effect is further accentuated because the negatively
charged $\Sigma^-$ and $\Xi^-$ can replace electrons in maintaining 
charge neutrality. The resulting drop in $\mu_e$ would mean a smaller value
of $\omega_{K^-}$ is required for $K^-$ condensation which leads to higher
threshold density. However, we find that in presence of hyperons the rate of 
decrease of the isovector potential is not so severe compared to that of
$\mu_e$. Therefore, in a hyperonic medium the shift in the threshold density 
for $\bar K^0$ condensate is much smaller than that of $K^-$. For example,
in the GM1 set, the critical densities are $u_{\rm cr} (K^-)=6.00n_0$
and $u_{\rm cr} (\bar K^0) = 6.61n_0$ for $U_{\bar K}(n_0)=-120$ MeV.
(This is to be compared with the results of Table III for stars with
nucleons and antikaons.)  These values are however beyond the central density
of $5.35n_0$ for maximum mass star containing all baryons but no antikaons.
On the other hand, for $U_{\bar K}(n_0)=-140$ MeV, the critical 
densities of $u_{\rm cr}(K^-)=3.41n_0$ and $u_{\rm cr}(\bar K^0)=4.42n_0$
imply that both $K^-$ and $\bar K^0$ condensates will be present in 
hyperonic stars. This means that even though hyperons shift the threshold
densities of $K^-$ and $\bar K^0$ condensates to higher values, the possibility
of finding hyperon stars with $\bar K^0$ condensate (in association 
with $K^-$) is enhanced for large $\bar K$ optical potential depths.

The standard mean field model used in the hyperon sector was found
\cite{Sch94} to be inadequate to describe the strongly attractive
hyperon-hyperon interaction observed in double $\Lambda$-hypernuclei.
Since the core of a neutron star can be hyperon-rich \cite{Gle1}, the
hyperon-hyperon interaction is expected to be important. This may be
accounted by considering two additional mesons, the scalar meson
$f_0(975)$ (denoted by $\sigma^*$ hereafter) and the vector meson
$\phi(1020)$ which couple only to the hyperons ($Y$) \cite{Sch,Sch94}. 
The corresponding Lagrangian is
\begin{eqnarray}
{\cal L}^{YY} &=& \sum_B {\overline \psi}_B 
\left( g_{\sigma^* B}\sigma^* -  g_{\phi B}\gamma_\mu\phi^\mu \right) 
\psi_B  \nonumber \\
&& + \frac{1}{2}\left(\partial_\mu\sigma^*\partial^\mu\sigma^* 
- m_{\sigma^*}^2\sigma^{*2} \right) - \frac{1}{4} \phi_{\mu\nu}\phi^{\mu\nu}
+ \frac{1}{2}m_\phi^2\phi_\mu\phi^\mu  ~ .
\end{eqnarray}
The $\phi-Y$ couplings are obtained from the SU(6) relation
\begin{equation}
2g_{\phi \Lambda} = 2g_{\phi \Sigma} = g_{\phi \Xi} 
= \frac{2\sqrt{2}}{3}g_{\omega N} ~.
\end{equation}
The $\sigma^*-Y$ couplings are obtained by fitting them to a well-depth
$U_Y^{(Y')}$ for a $Y$ in a $Y'$-bath \cite{Sch,Sch94}. Note that the
nucleons do not couple to the strange mesons, i.e.
$g_{\sigma^* N} = g_{\phi N} = 0$.  The EOS and the star sequence are 
shown for this situation in Figs. 5 and 6 (dotted lines) for
the GM1 set without antikaons. The additional attraction imparted by
the $\sigma^*$ field makes the EOS softer at moderately high densities.
At very high densities, the EOS merges with that without the strange
mesons due to dominance of the repulsive $\phi$ field over the
$\sigma^*$ field. The maximum star mass obtained here is $1.721M_\odot$
at a central density of $5n_0$.

The rather large values of the meson fields ($\sigma^*,\phi$) in a
hyperonic star should influence antikaon condensation densities.
From the strangeness content of $\bar K$ ($s$-quark content) it is
evident that the $\sigma^*$ field is attractive and the $\phi$ field
is repulsive for antikaons as also for the hyperons. As in Ref. 
\cite{Sch}, the $\phi-K$ coupling is obtained from the SU(3) relation,
$g_{\phi K} = 6.04/\sqrt{2}$, and the $\sigma^*-K$ coupling from the
$f_0(975)$ decay, $g_{\sigma^* K} = 2.65$. With these couplings and
for $U_{\bar K}(n_0)=-140$ MeV, we find that the relatively strong 
repulsive $\phi$ field in a hyperon-rich medium prevents antikaon 
condensation even at densities as large as $7.40n_0$ (in the GM1 set) where 
the effective nucleon mass $m_N^*$ drops to zero \cite{Kno,Sch}. Only for
large $\bar K$ optical depth of $-160$ MeV, and thereby large 
$g_{\sigma K}$ value, critical densities of $u_{\rm cr}(K^-)=2.76n_0$ 
and $u_{\rm cr}(\bar K^0)=4.13n_0$ are reached.

The above discussions of hyperonic star matter are confined to
$\Sigma$-depth of $-30$ MeV. Repulsive depth of 
$U^N_{\Sigma}(n_0)\approx + 30$ MeV has been also estimated \cite{Fri94}.
On refitting the scalar coupling $g_{\sigma \Sigma}$ to this depth and
limiting to the Lagrangian of Eq. (1) only, we find that in all the
models $\Sigma^-$ formation occurs at a much higher density and the
resulting EOS is now  stiffer compared to those with attractive
$\Sigma$-depth. For example, in the GM1 set, a star of maximum mass 
$1.798M_\odot$ is formed with a central density of $5.13n_0$. All the
$\Sigma$ particles do not exist in this star while the $\Xi^-$ occurs
at a early density of $2.6n_0$.

In this paper, we have employed the relativistic mean field model where 
the nucleon-meson coupling constants are fitted to certain bulk 
properties at normal nuclear matter density. The model is then extended 
to high density regime to investigate the effects of antikaon condensation. 
The underlying assumption in the RMF models is that the meson field operators
in the Euler-Lagrange equations are replaced by their mean values because
the source terms in those equations increase with baryon density. Also, the
coupling constants of the RMF models are density independent. There are 
other approaches where gross properties of dense matter relevant to
neutron stars are calculated in the framework of nonrelativistic 
Brueckner-Hartree-Fock and variational chain summation (VCS) calculations 
using modern nucleon-nucleon interaction \cite{Akm,Pet,Hei}. Recently, 
Akmal, Pandharipande and Ravenhall (APR) \cite{Akm} have studied neutron star 
properties in the VCS method using a new Argonne $V_{18}$ nucleon-nucleon
interaction (which fit all the data of the Nijmegen data base) with 
relativistic boost corrections and a fitted three-nucleon interaction.
This calculation represents at present perhaps the most sophisticated
many-body approach to dense matter. The bulk properties obtained in the
APR model are binding energy $E/B = 16$ MeV, nuclear symmetry energy 
$a_{\rm asy} = 33.94$ MeV, and incompressibility $K = 266$ MeV at a
saturation density of $n_0 = 0.16$ fm$^{-3}$. To gauge 
the uncertainties involved in the extrapolation of the RMF model at high 
density, we compare the nucleons-only matter results with those of the 
APR model. For this purpose we use the model which contains nonlinear
interaction for the scalar $\sigma$-meson only. (Note that the TM1 model 
which includes also self-interaction of $\omega$-meson is fitted to finite 
nuclear properties.) The five parameters in this RMF model are fitted to the 
above mentioned (four) saturation properties plus the effective nucleon mass of
$m^*_N/m_N = 0.70$ and 0.78 for the two representative cases considered here.
In Fig. 13 we show the results for pure neutron matter (PNM) and symmetric
nuclear matter (SNM) in the RMF model (solid lines), and in the APR model 
with pion condensation (dashed lines) and their low density extrapolations 
without a pion condensed phase (dotted lines) \cite{Akm}. 
It is clear from this figure  
that the agreement between the two models breaks down at high density. 
The APR model takes into account possibly all leading many-body correlation
effects. It was also demonstrated \cite{Akm,Pet} that the strong tensor 
correlation gives rise to neutral pion condensation in PNM and SNM at
densities of 0.20 fm$^{-3}$ and 0.32 fm$^{-3}$, respectively. The RMF
model, however, cannot address the question of density dependent correlation 
in dense matter, and moreover the coupling constants are density independent. 
This may be crucial to the deviation of the RMF results from the APR 
calculations. Consequently, the threshold densities of antikaon condensation 
may be affected. On the other hand, the advantage of the RMF models is
that relativity is in-built and this causes the EOS to be causal.
However, in the APR model the EOS becomes superluminal at densities 
close to the maximum masses of the stars.

We have computed nucleons-only star masses using this parameter set in the 
RMF model. For the case $m^*_N/m_N = 0.70$, the maximum masses of the neutron 
star matter (pure neutron matter) are $M_{\rm max}=2.29$ $(2.45)M_{\odot}$ 
at central densities of $n_{\rm cent}=0.92$ (0.80) fm$^{-3}$. For the 
case $m^*_N/m_N=0.78$, the respective values are
$M_{\rm max}=2.00$ $(2.20)M_{\odot}$ at $n_{\rm cent}=1.09$ (0.91) fm$^{-3}$. 
Comparing with the respective results of $M_{\rm max}= 2.20$ $(2.21)M_\odot$ 
in the APR model with pion condensation, we find that in the RMF model the 
maximum mass of a neutron star is more sensitive to the amount of proton 
fraction. It was shown \cite{Lat} that a neutron star may cool rapidly by the
so-called direct URCA process in which neutrinos generated in the
stars interior carry away the energy. The threshold density for this process
depends on the symmetry energy which in turn is determined by 
the proton fraction. For the APR model the threshold is at a density
of $n_B=0.78$ fm$^{-3}$ corresponding to a star of mass $\approx 2.0M_\odot$.
In the RMF model, for $m^*_N/m_N = 0.70$ and 0.78, the threshold densities
are respectively, 0.59 and 0.61 fm$^{-3}$, which correspond to neutron stars  
of masses 2.09 and $1.78M_\odot$. The relatively rapid rise of symmetry
energy with density manifests in a smaller threshold density for the direct
URCA process in the RMF models.

It may be worth mentioning that the role of nucleon-nucleon and kaon-nucleon
correlations in kaon condensation in neutron star matter was studied by
Pandharipande et al. \cite{Pan}. The kaon energy
for low density matter was calculated from Lenz potential while at high
density the Hartree potential was used. The strong $NN$ and $KN$ correlations
in the Hartree potential were shown to lead to a dramatic reduction of the
antikaon attraction at high densities that could raise the threshold
density for antikaon condensation to much higher values. Recently, 
kaon energy in neutron matter has been calculated analytically \cite{Car} 
by solving the Klein-Gordon equation in the Wigner-Seitz cell approximation.
It was found that the transition from the low density Lenz potential to 
a high density Hartree potential occurs at $\sim 4 n_0$. However, in a chiral
perturbation theory, Waas et al. \cite{Waa2} found that short-range 
correlations has only a moderate effect on the antikaon condensation densities.

\section{Summary and Conclusions}

We have investigated antikaon, $K^-$- and $\bar K^0$-meson, condensation in
neutron stars within a relativistic mean field approach where the interaction
between the baryons and antikaons are generated by the exchange of $\sigma$,
$\omega$, and $\rho$ mesons. Three different parameter sets (GM1, TM1, and
GMT) have been exploited in this calculation. In the GM1 and GMT sets 
the Lagrangian contain self-interaction for $\sigma$-meson only, while the
set TM1 incorporates also nonlinear interaction for $\omega$-meson. The
softest nuclear equation of state follows from TM1 set whereas the stiffest
one from GMT set. We find that the critical densities for 
$K^-$ and in particular $\bar K^0$ condensations depend 
sensitively on the choice of antikaon optical potential depth and more 
strongly on the nuclear equation of state. The threshold density of $\bar K^0$
always lie above that of $K^-$ meson. With the appearance of $K^-$ and 
$\bar K^0$ condensates, the overall equation of state becomes softer compared
to the situation without antikaon condensation. This leads to a reduction in 
the maximum star masses. With the softest EOS (TM1 set), $\bar K^0$ condensate
may be formed inside maximum mass stars only when the $\bar K$ optical
depth is quite large. On the other hand for the stiffest nuclear EOS (GMT set),
$\bar K^0$ can occur well inside the maximum mass stars for rather small 
values of antikaon optical potential depth. The appearance of $K^-$ meson leads
to a smaller variation of the vector fields with density. The decrease of
$\bar K^0$ energy with density is therefore attenuated resulting in its 
formation at higher density than in the case when $K^-$ meson is prohibited.

With the onset of $K^-$ condensation (neglecting $\bar K^0$ meson), the
proton fraction rises dramatically and even crosses the neutron fraction
at some critical density. From the threshold conditions for $K^-$ and
$\bar K^0$ formations, i.e. $\omega_{K^-} =\mu_n -\mu_p$
and $\omega_{\bar K^0} =0$, it is evident that the critical point of
identical proton and neutron densities (Fermi momenta) is precisely the
density where $\bar K^0$ formation sets in. In earlier model studies 
of $K^-$ condensation only (where $\bar K^0$ condensation were ignored),
identical values of neutron and proton abundances were observed at high
densities after $K^-$ condensation. It may therefore be instructive to revisit 
the antikaon condensation scenario including $\bar K^0$ meson in these models.
With the onset of $\bar K^0$ condensation, there is a competition in the
formation of ${K^-}-p$ and ${\bar K^0}-n$ pairs resulting in a perfectly
symmetric matter for nucleons and antikaons inside neutron stars. As the
proton fraction is much enhanced in this situation, it may have profound 
implication on the cooling properties of neutron stars.

With the hyperon-meson couplings constants determined from hypernuclear
data and $SU(6)$ symmetry relation, the hyperons are formed earlier than
antikaons. Since the EOS becomes softer, antikaon condensation can only
occur at high densities. The negatively charged hyperons cause a larger
reduction in the electron chemical potential compared to that of the isovector
potential. This indicates that in contrast to nucleons only stars, $\bar K^0$
has a higher formation probability in hyperonic stars, but only for
relatively larger values of $\bar K$ optical potential depth.

For all the cases studied here, the maximum masses of the stars are found to
be larger than the precise current observational lower limit of $1.44 M_\odot$
imposed by the larger mass of the binary pulsar PSR 1913 + 16 \cite{Wei}.
The recent discovery of high-frequency brightness oscillations in low-mass
X-ray binaries provides a promising new avenue for determining masses and
radii of neutron stars \cite{Mil}. Large neutron star masses with an upper
limit of $M \sim 2.0-2.3 M_\odot$ have been extracted from kilohertz 
quasi-periodic-oscillations (QPO). The present investigation 
suggests that indeed both $K^-$ and
$\bar K^0$ condensates are possible in stars of such large masses. On the
other hand, hyperon degrees of freedom which lower the maximum mass
considerably, would be clearly excluded in these massive stars. In contrast,
also very small masses of $M \approx 1.35 \pm 0.04M_\odot$ have been
accurately determined in binary pulsars \cite{Tho99}. It is clear that 
exotic degrees of freedom like antikaons and hyperons must occur in these 
pulsars at few times the normal nuclear matter density.

It has been inferred that large magnetic field $\sim 10^{18}$G could exist
at the core of neutron stars \cite{Cha}. Though such large interior field is
not accessible to direct observation, their existence could influence certain
gross properties of neutron stars, such as cooling properties of the stars
\cite{Ban}. Recently, there has been a considerable debate about whether a
charged Bose gas in a constant magnetic field could exhibit a 
Bose-Einstein condensation \cite{Scha,Elm,Roj,Suh}. Some 
calculations ruled out such a possibility \cite{Scha,Elm} whereas others 
predicted that it could occur in a magnetic field \cite{Roj,Suh}.
The formation of a Bose-Einstein condensate of neutral Bose gas such as a 
system of $\bar K^0$ in a strong magnetic field may be an interesting 
possibility.

\acknowledgments

We would like to thank J\"urgen Schaffner-Bielich and Avraham Gal for
helpful discussions and remarks. S.P. was financially supported by the 
Alexander von Humboldt Foundation.



\begin{table}

\caption{The nucleon-meson coupling constants in relativistic mean
field models. The parameter set GM1 is taken from Ref. [20] where the baryons
interact via nonlinear $\sigma$-meson and linear $\omega$-meson exchanges. 
The coupling constants are obtained by reproducing the nuclear matter 
properties of binding energy $E/B=-16.3$ MeV, baryon density $n_0=0.153$ 
fm$^{-3}$, asymmetry energy coefficient $a_{\rm asy}=32.5$ MeV, 
incompressibility $K=300$ MeV, and effective nucleon mass $m^*_N/m_N = 0.70$. 
The hadronic masses are $m_N=938$ MeV, $m_\sigma=550$ MeV, 
$m_\omega=783$ MeV, and $m_\rho=770$ MeV. The parameter set
TM1 is obtained from Ref. [35] which has nonlinear exchanges in 
both $\sigma$ and $\omega$ mesons. The nuclear matter properties in TM1 set
are $E/B=-16.3$ MeV, $n_0=0.145$ fm$^{-3}$, $a_{\rm asy}=36.9$ MeV, 
$K=281$ MeV, and $m^*_N/m_N = 0.634$. All the hadronic masses in 
this model are same as GM1 except for $\sigma$-meson which is 
$m_\sigma=511.198$ MeV. The parameter set GMT corresponds to nonlinear 
$\sigma$ and linear $\omega$-meson exchanges as in GM1, however,
the nuclear matter saturation properties and $\sigma$-meson mass of 
the TM1 set are used. All the parameters are dimensionless, except $g_2$ 
which is in fm$^{-1}$.}

\begin{tabular}{ccccccc} 

\hfil& $g_{\sigma N}$& $g_{\omega N}$& $g_{\rho N}$&
$g_2$& $g_3$& $g_4$ \\ \hline
GM1& 9.5708& 10.5964& 8.1957& 12.2817& -8.9780& $-$ \\
TM1& 10.0289& 12.6139& 4.6322& -7.2325& 0.6183& 71.3075 \\
GMT& 9.9400& 12.2981& 9.2756& 10.5745& -24.1907& $-$ 

\end{tabular}
\end{table}
\vspace{1.5cm}

\begin{table}

\caption{The couplings constants for the antikaons $\bar K$ to the 
$\sigma$-meson, $g_{\sigma K}$, for various values of $\bar K$ optical 
potential depths $U_{\bar K}$ (in MeV) at the saturation density. The 
results are for the GM1, TM1, and GMT sets. The couplings constants of 
$\bar K$ to the vector $\omega$ and $\rho$ mesons are fixed by $SU(6)$ 
relations.}

\begin{tabular}{cccccc} 

$U_{\bar K}$& -100& -120& -140& -160& -180 \\ \hline
GM1& 0.9542& 1.6337& 2.3142& 2.9937& 3.6742\\
TM1& 0.2537& 0.8384& 1.4241& 2.0098& 2.5955 \\
GMT& 0.2435& 0.8220& 1.4015& 1.9800& 2.5596

\end{tabular}
\end{table}
\vspace{1.5cm}

\begin{table}
\caption{The maximum masses $M_{\rm max}$ and the corresponding central 
densities $u_{\rm cent}=n_{\rm cent}/n_0$ for nucleons-only 
($np$) star matter and for stars with further inclusion of hyperons ($npH$). 
The results are for the GM1 and TM1 set. For neutron star matter with nucleons 
and antikaons ($np\bar K$), the critical densities for the $K^-$ and $\bar K^0$
condensation, $u_{\rm cr}(K^-)$ and $u_{\rm cr}(\bar K^0)$, and results for
$M_{\rm max}$ and $u_{\rm cent}$ at various values of $\bar K$ optical 
depth $U_{\bar K}$ (in MeV) at saturation density are also given.} 

\begin{tabular}{cccccc|cccc} 

\hfil& \hfil& \hfil& GM1& \hfil& \hfil& \hfil& TM1&
\hfil& \hfil \\ \hline
\hfil& $U_{\bar K}$& $u_{\rm cr}(K^-)$& $u_{\rm cr}(\bar K^0)$& $u_{\rm cent}$&
$M_{\rm max}/M_\odot$& $u_{\rm cr}(K^-)$& $u_{\rm cr}(\bar K^0)$& 
$u_{\rm cent}$& $M_{\rm max}/M_\odot$ \\ \hline
$np$& $-$& $-$& $-$& 5.63& 2.364& $-$& $-$& 5.97& 2.179 \\
\hfil& -100& 3.45& 5.51& 5.17& 2.211& 4.15& 11.12& 5.67& 2.142 \\
\hfil& -120& 3.05& 4.83& 5.19& 2.077& 3.55& 9.13& 5.55& 2.083 \\
$npK$& -140& 2.71& 4.19& 4.75& 1.856& 3.05& 7.43& 5.65& 1.986 \\
\hfil& -160& 2.43& 3.59& 3.59& 1.551& 2.67& 5.99& 6.37& 1.857 \\
\hfil& -180& 2.19& 3.07& 3.09& 1.217& 2.37& 4.81& 6.01& 1.641 \\
$npH$& $-$& $-$& $-$& 5.35& 1.784& $-$& $-$& 5.87& 1.653 

\end{tabular}
\end{table}
\vspace{1.5cm}

\begin{table}
\caption{Same as in Table III, but for the GMT set. The maximum masses
and corresponding central densities for stars with $K^-$ condensation only 
without any $\bar K^0$ formation are given in the parentheses.}

\begin{tabular}{cccccc} 

\hfil& $U_{\bar K}$& $u_{\rm cr}(K^-)$& $u_{\rm cr}(\bar K^0)$& 
$u_{\rm cent}$& $M_{\rm max}/M_\odot$ \\ \hline

$np$& $-$& $-$& $-$& 4.91& 2.667 \\
\hfil& -100& 3.23& 4.95& 4.69& 2.546 \\
\hfil& -120& 2.87& 4.45& 4.77(4.76)& 2.444(2.446) \\
$npK$& -140& 2.57& 3.97& 4.75(5.17)& 2.279(2.309)\\
\hfil& -160& 2.31& 3.51& 4.78(5.52)& 2.012(2.162) \\
\hfil& -180& 2.09& 3.05& 4.37(4.67)& 1.584(1.929)

\end{tabular}
\end{table}
\vspace{1.5cm}

\begin{table}
\caption{The critical densities for $\bar K^0$ meson condensation,
$u_{\rm cr}(\bar K^0)$ (where $u=n_B/n_0$), for various values of $\bar K$ 
optical potential depths $U_{\bar K}(n_0)$ (in MeV) in star matter with 
nucleons 
and $\bar K^0$ mesons only, neglecting $K^-$ formation. The maximum 
masses of the stars $M_{\rm max}$ and the corresponding central densities 
$u_{\rm cent}$ are given. The results are for the GM1, TM1, and GMT 
parameter sets.}

\begin{tabular}{cccc|ccc|ccc} 

\hfil& \hfil& GM1& \hfil& \hfil& TM1& \hfil& \hfil&
GMT& \hfil \\ \hline
$U_{\bar K}$& $u_{\rm cr}(\bar K^0)$& $u_{\rm cent}$&
$M_{\rm max}/M_\odot$& $u_{\rm cr}(\bar K^0)$& $u_{\rm cent}$&
$M_{\rm max}/M_\odot$& $u_{\rm cr}(\bar K^0)$& $u_{\rm cent}$&
$M_{\rm max}/M_\odot$ \\ \hline

-100& 4.80& 5.57& 2.349& 9.35& 5.97& 2.179& 4.32& 4.94& 2.655 \\
-120& 4.27& 5.35& 2.303& 7.70& 5.97& 2.179& 3.91& 4.80& 2.623 \\
-140& 3.76& 5.02& 2.205& 6.32& 5.97& 2.179& 3.48& 4.64& 2.553 \\
-160& 3.32& 4.85& 2.031& 5.22& 6.04& 2.170& 3.13& 4.41& 2.423 \\
-180& 2.95& 5.35& 1.700& 4.33& 5.73& 2.105& 2.79& 4.08& 2.194 

\end{tabular}
\end{table}

 \newpage 
\vspace{-2cm}

{\centerline{
\epsfxsize=14cm
\epsfysize=17cm
\epsffile{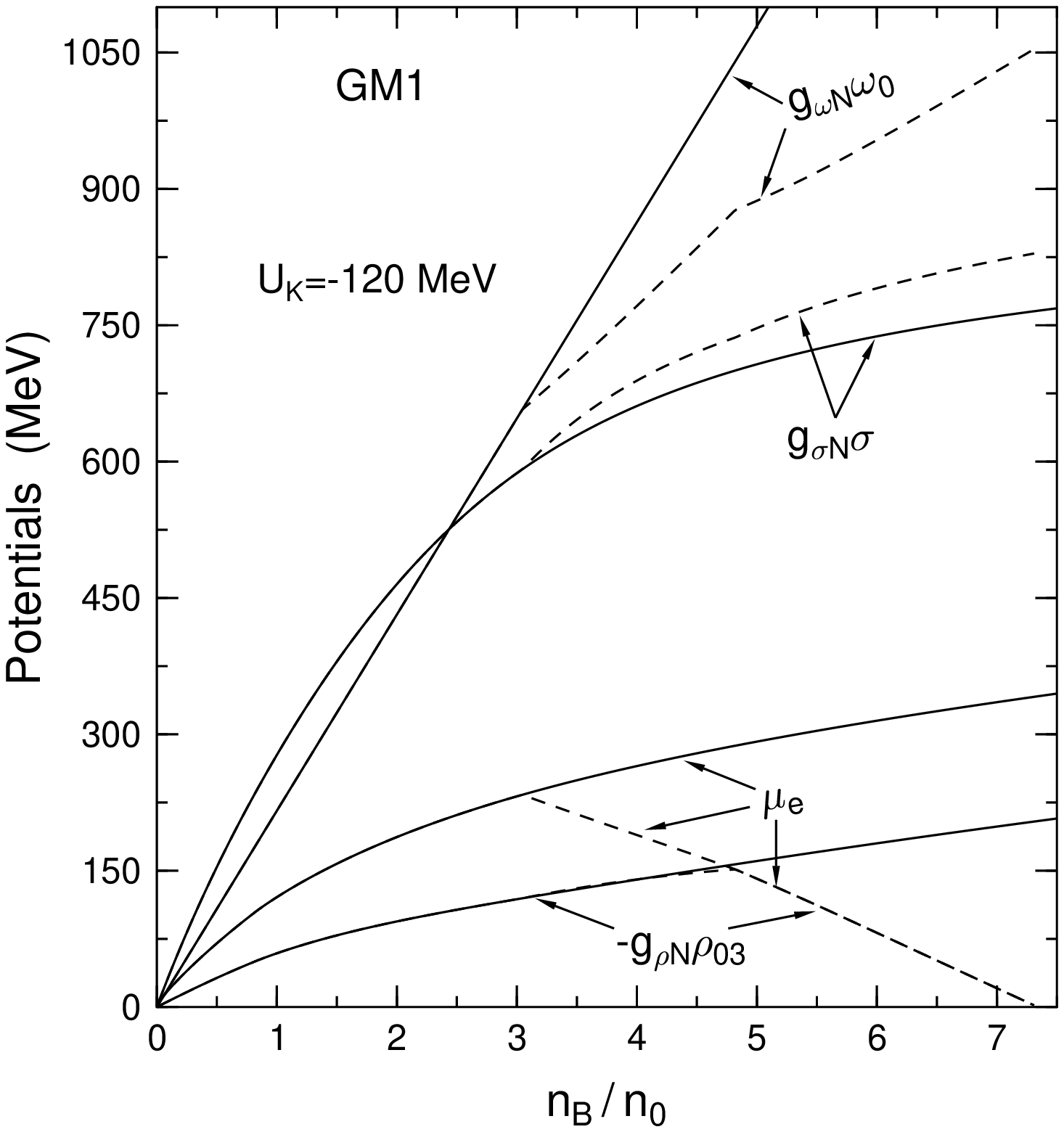}
}}

\vspace{-5cm}

\noindent{\small{
FIG. 1. The mean meson potentials for nucleons and the electrochemical 
potential versus the baryon density $n_B/n_0$ in the GM1 set for 
nucleons-only star matter (solid lines) and for matter with further
inclusion of antikaon, $K^-$ and $\bar K^0$, condensation (dashed lines).
The $\bar K$ optical potential depth is $U_{\bar K}=-120$ MeV at the normal 
nuclear matter density of $n_0=0.153$ fm$^{-3}$. Above the critical density for
$\bar K^0$ condensation, $n_{\rm cr}(\bar K^0) \geq 4.83 n_0$, the 
electrochemical potential $\mu_e$ coincides with 
$-V_\rho^N = -g_{\rho N}\rho_{03}$.}}
 \newpage 
\vspace{-2cm}

{\centerline{
\epsfxsize=14cm
\epsfysize=22cm
\epsffile{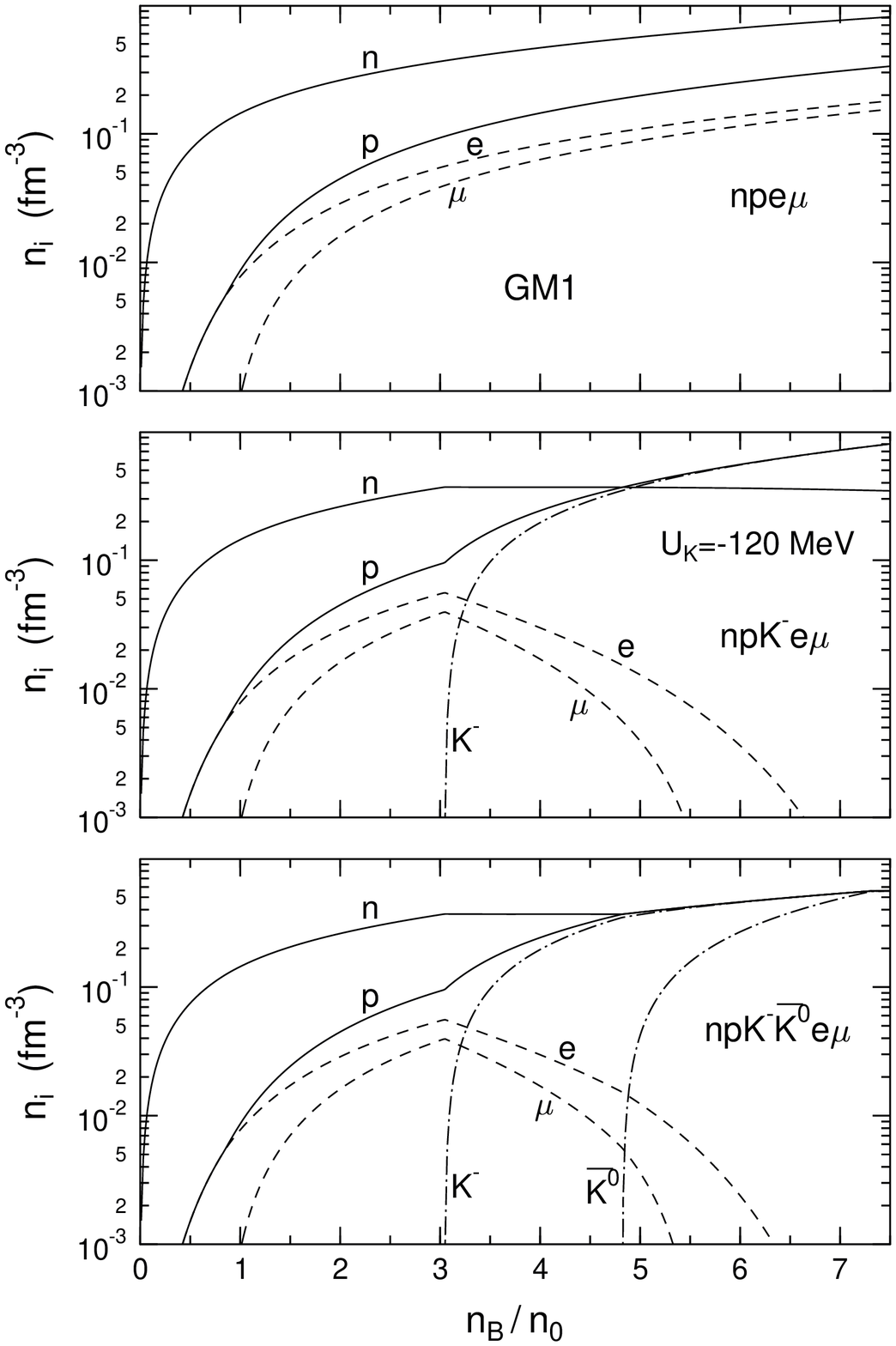}
}}

\vspace{-1.4cm}

\noindent{\small{
FIG. 2. The proper number densities $n_i$ of various compositions
in neutron star matter in the GM1 model. The results are for 
nucleons-only matter (top panel), matter with further inclusion of $K^-$ 
condensation (central panel), and matter where also $\bar K^0$ condensation 
is considered (bottom panel). The $\bar K$ optical potential at normal 
nuclear matter density is $U_{\bar K} = -120$ MeV.}}
 \newpage 
\vspace{-2cm}

{\centerline{
\epsfxsize=14cm
\epsfysize=17cm
\epsffile{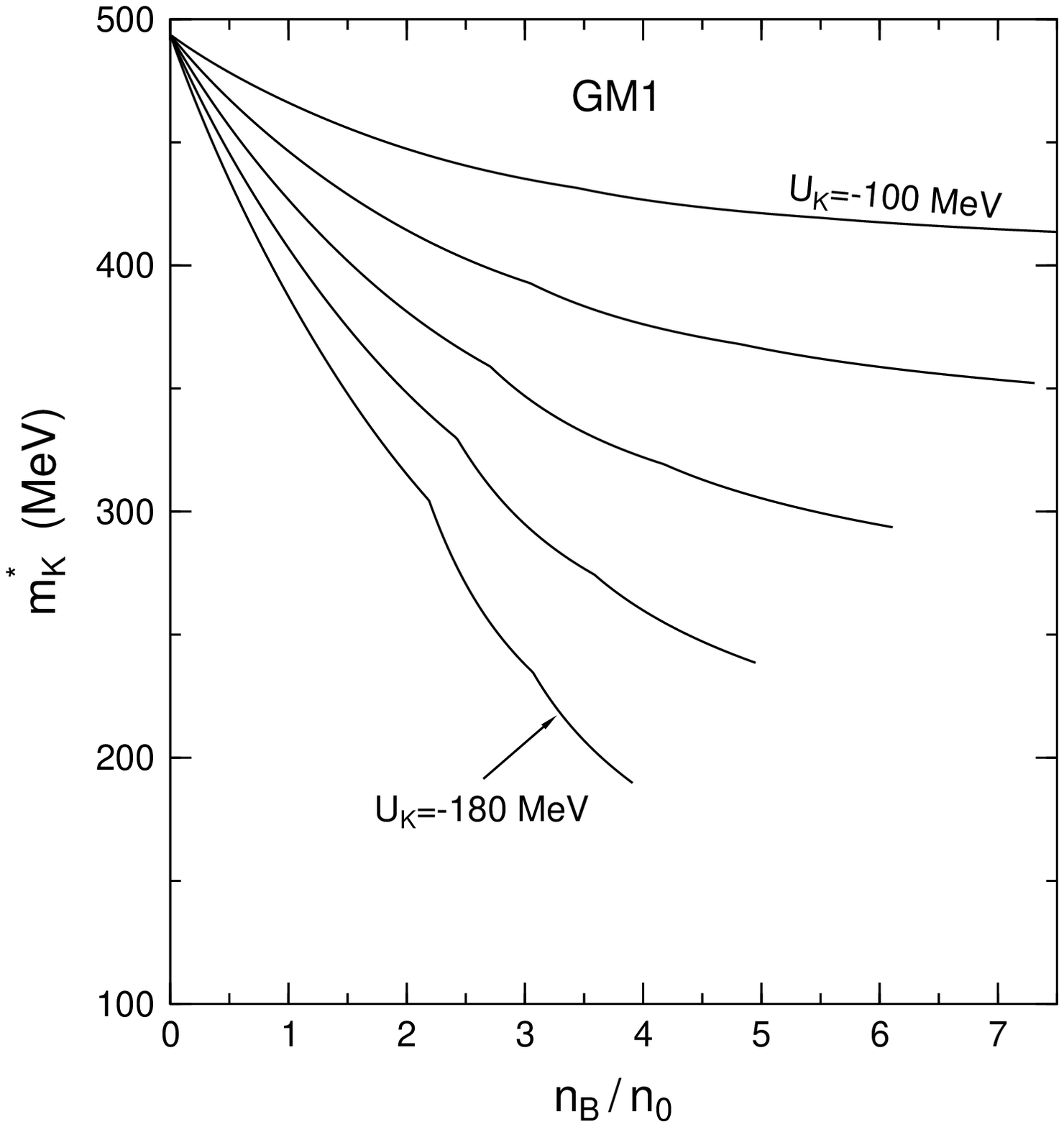}
}}

\vspace{-5cm}

\noindent{\small{
FIG. 3. The variation of effective antikaon mass $m_K^*/m_K$ as a function
of baryon density $n_B/n_0$ for star matter with nucleons and antikaons in 
the GM1 set. The different curves from the top to bottom on the left side
of the graph correspond to $\bar K$ optical potential depths at normal nuclear
matter density of $U_{\bar K}= -100, -120, -140, -160, -180$ MeV.}}
 \newpage 
\vspace{-2cm}

{\centerline{
\epsfxsize=14cm
\epsfysize=17cm
\epsffile{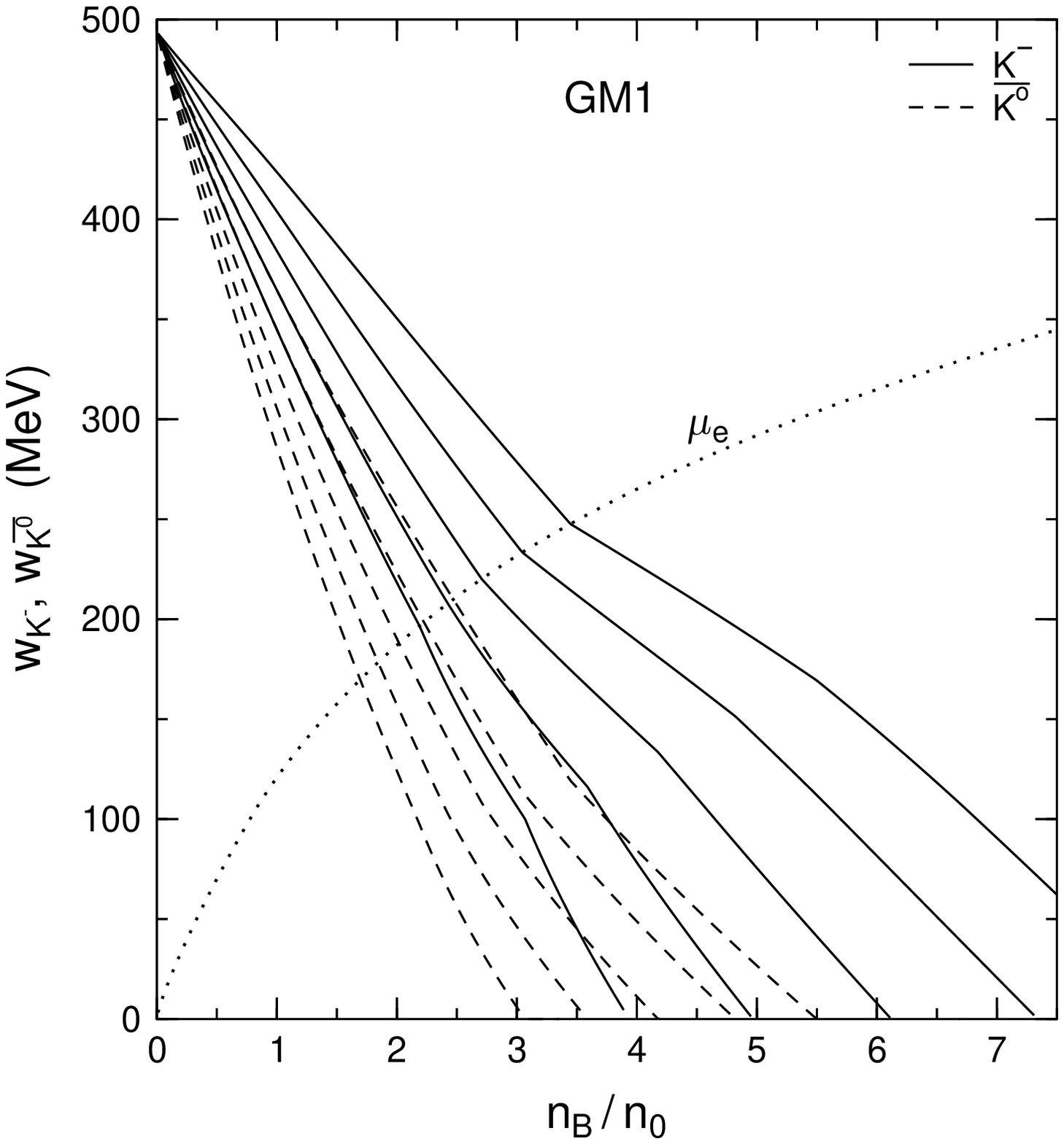}
}}

\vspace{-5cm}

\noindent{\small{
FIG. 4. The effective energy of $K^-$ (solid lines) and $\bar K^0$ 
(dashed lines) versus baryon density in the GM1 set. 
The different curves for each antikaon have the same meaning as in Fig. 3.}}
 \newpage 
\vspace{-2cm}

{\centerline{
\epsfxsize=14cm
\epsfysize=17cm
\epsffile{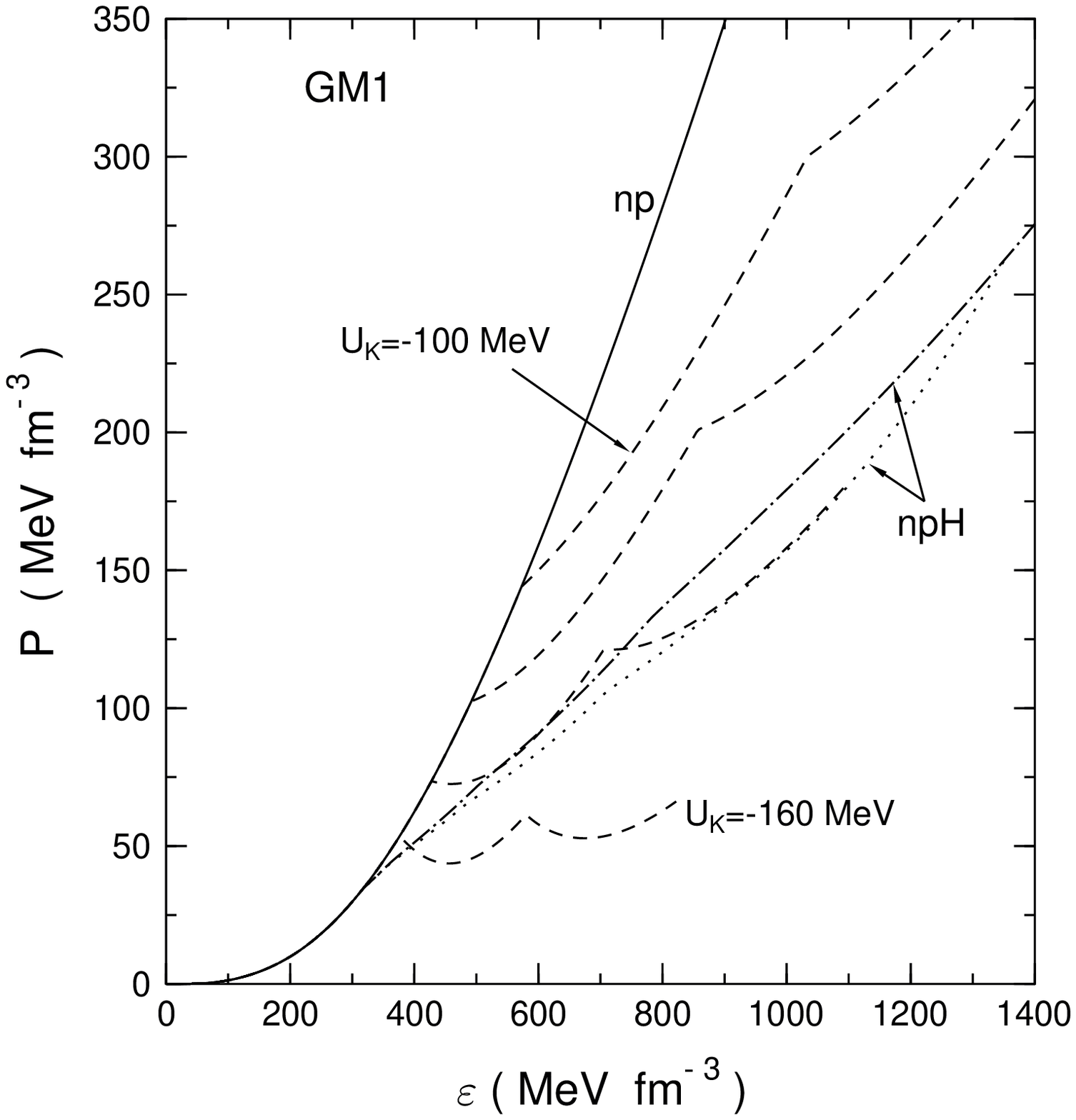}
}}

\vspace{-5cm}

\noindent{\small{
FIG. 5. The equation of state, pressure $P$ vs. energy density $\varepsilon$ 
in the GM1 set. The results are for nucleons-only ($np$) star matter 
(solid line) and for matter with further inclusion of $K^-$ and $\bar K^0$ 
condensation (dashed lines) for antikaon optical potential depths at normal 
density of $U_{\bar K}= -100, -120, -140, -160$ MeV. The equations of state 
for stars with nucleons and hyperons without any antikaons ($npH$) are 
shown for Lagrangian of Eq. (1) (dash-dotted line) and with further inclusion
of Eq. (29) (dotted line).}}
 \newpage 
\vspace{-2cm}

{\centerline{
\epsfxsize=14cm
\epsfysize=17cm
\epsffile{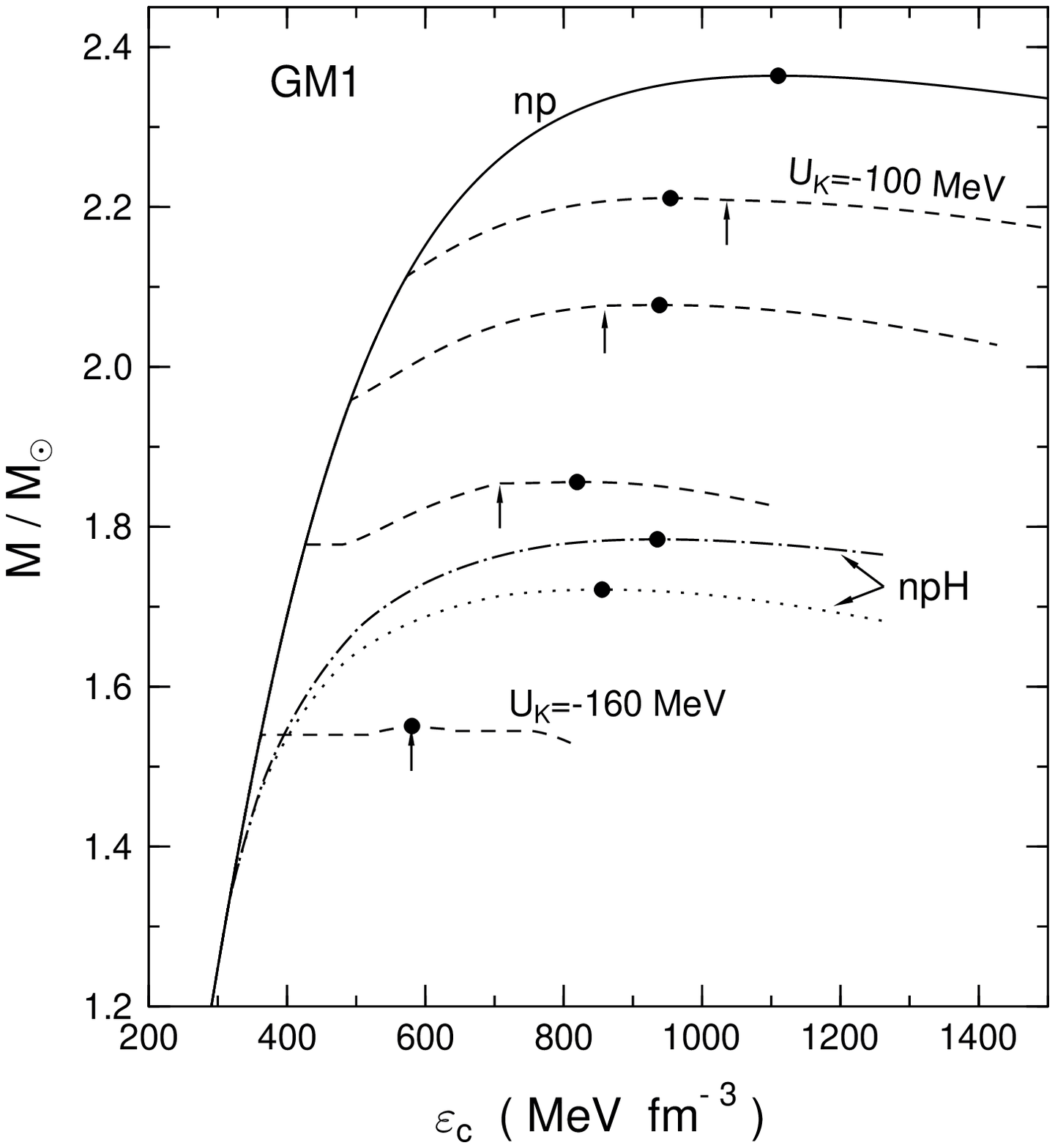}
}}

\vspace{-5cm}

\noindent{\small{
FIG. 6. The neutron star sequences near the limiting mass in the GM1 set.
The different curves have the same meaning as in Fig 5. The filled circles 
correspond to the maximum masses, and the arrows indicate the minimum
mass stars that possess a $\bar K^0$ condensate at their centers.}}
 \newpage 
\vspace{-2cm}

{\centerline{
\epsfxsize=14cm
\epsfysize=17cm
\epsffile{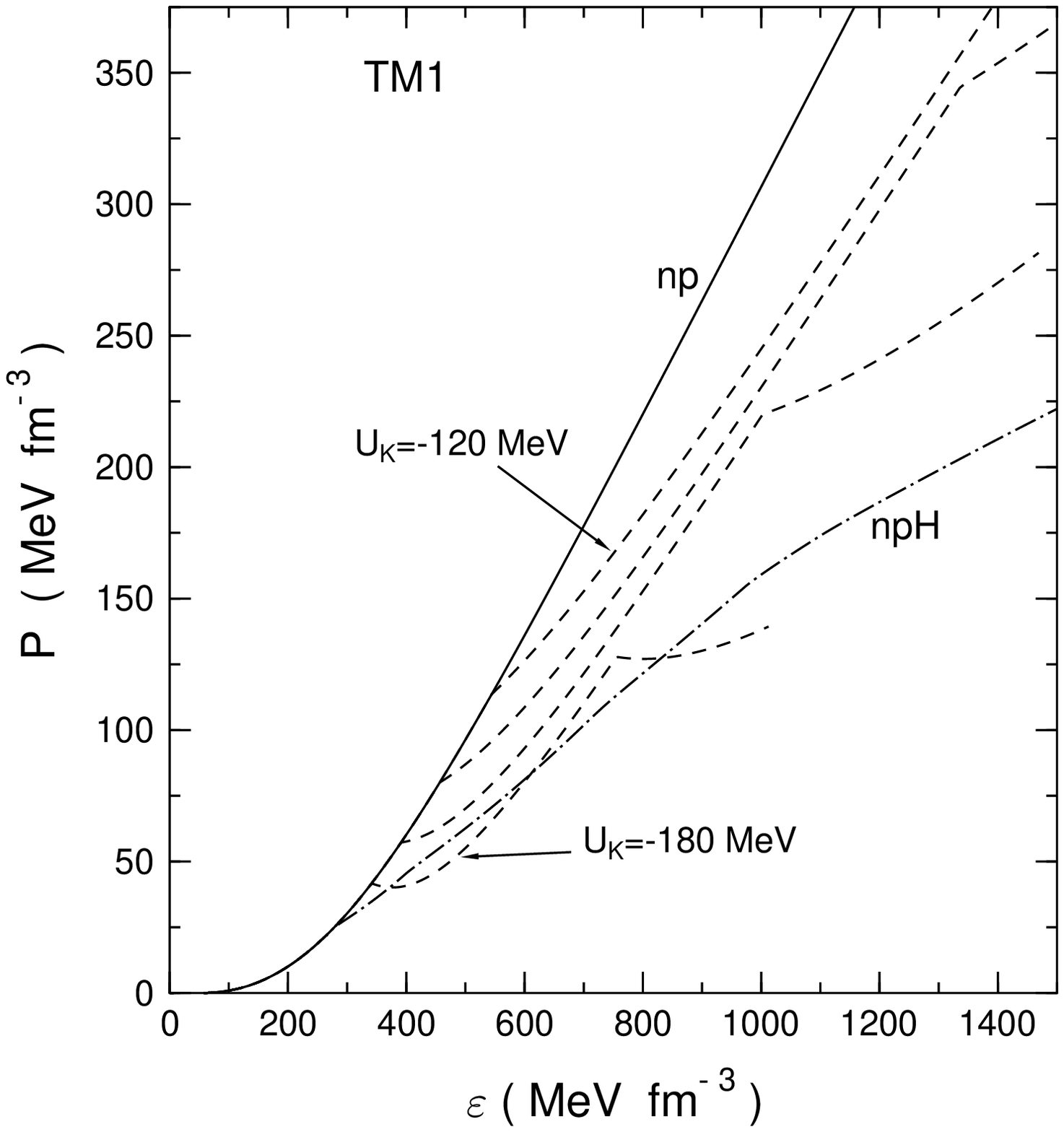}
}}

\vspace{-5cm}

\noindent{\small{
FIG. 7. Same as Fig. 5 but for the TM1 set. The antikaon optical potential 
depths at normal nuclear matter density are 
$U_{\bar K}= -120, -140, -160, -180$ MeV.}}
 \newpage 
\vspace{-2cm}

{\centerline{
\epsfxsize=14cm
\epsfysize=17cm
\epsffile{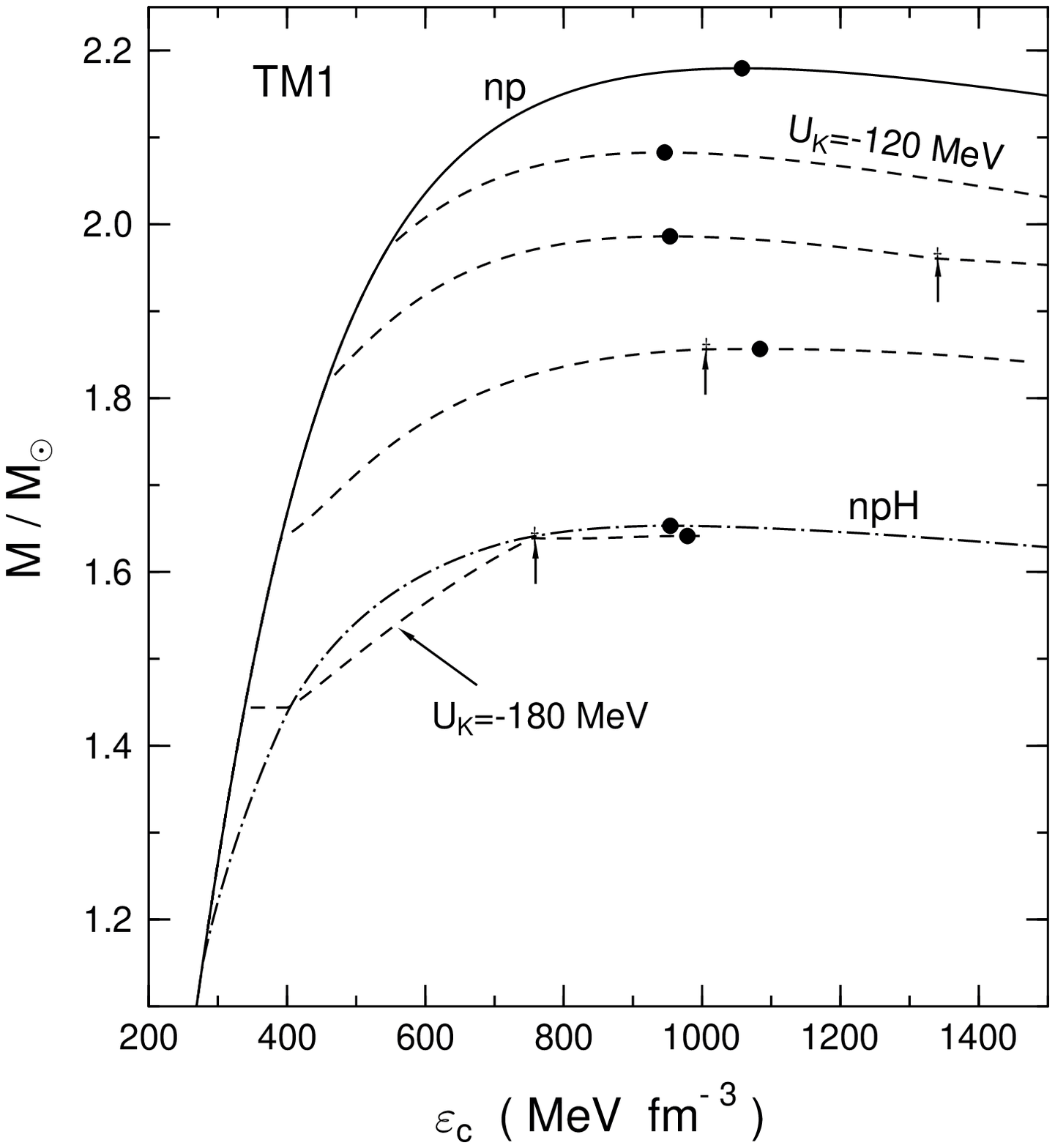}
}}

\vspace{-5cm}

\noindent{\small{
FIG. 8. Same as Fig. 6 but for the TM1 set. The antikaon optical potential 
depths are same as in Fig. 7.}}
 \newpage 
\vspace{-2cm}

{\centerline{
\epsfxsize=14cm
\epsfysize=20cm
\epsffile{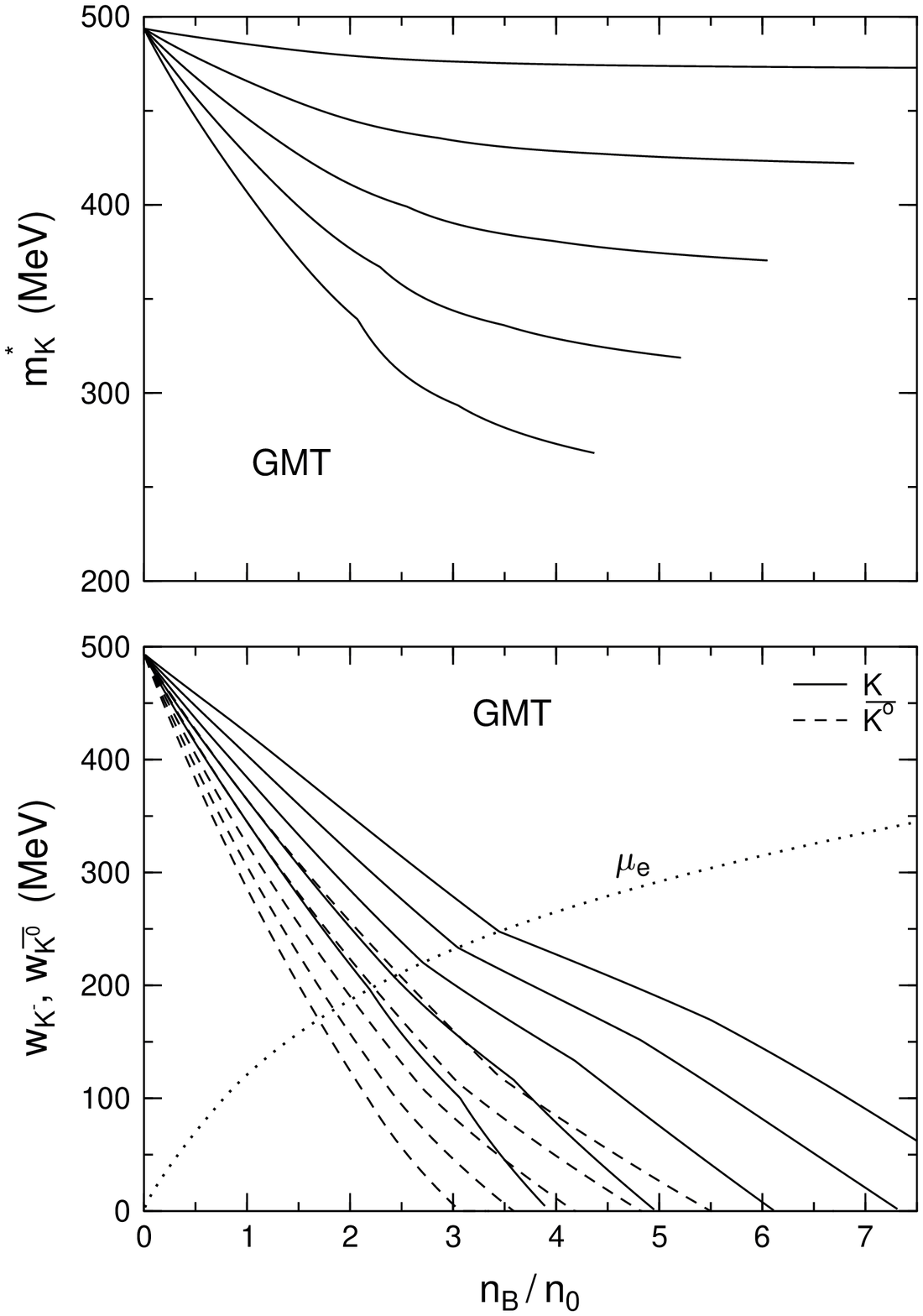}
}}

\vspace{-1.4cm}

\noindent{\small{
FIG. 9. In the top panel the effective antikaon mass $m_K^*/m_K$,
and in the bottom panel the effective energy of $K^-$ (solid lines) and 
$\bar K^0$ (dashed lines) versus baryon density in the GMT set. 
The different curves from the top to bottom on the left side of the graph
in each panel correspond to 
$U_{\bar K}(n_0)= -100, -120, -140, -160, -180$ MeV.}}
 \newpage 
\vspace{-2cm}

{\centerline{
\epsfxsize=14cm
\epsfysize=17cm
\epsffile{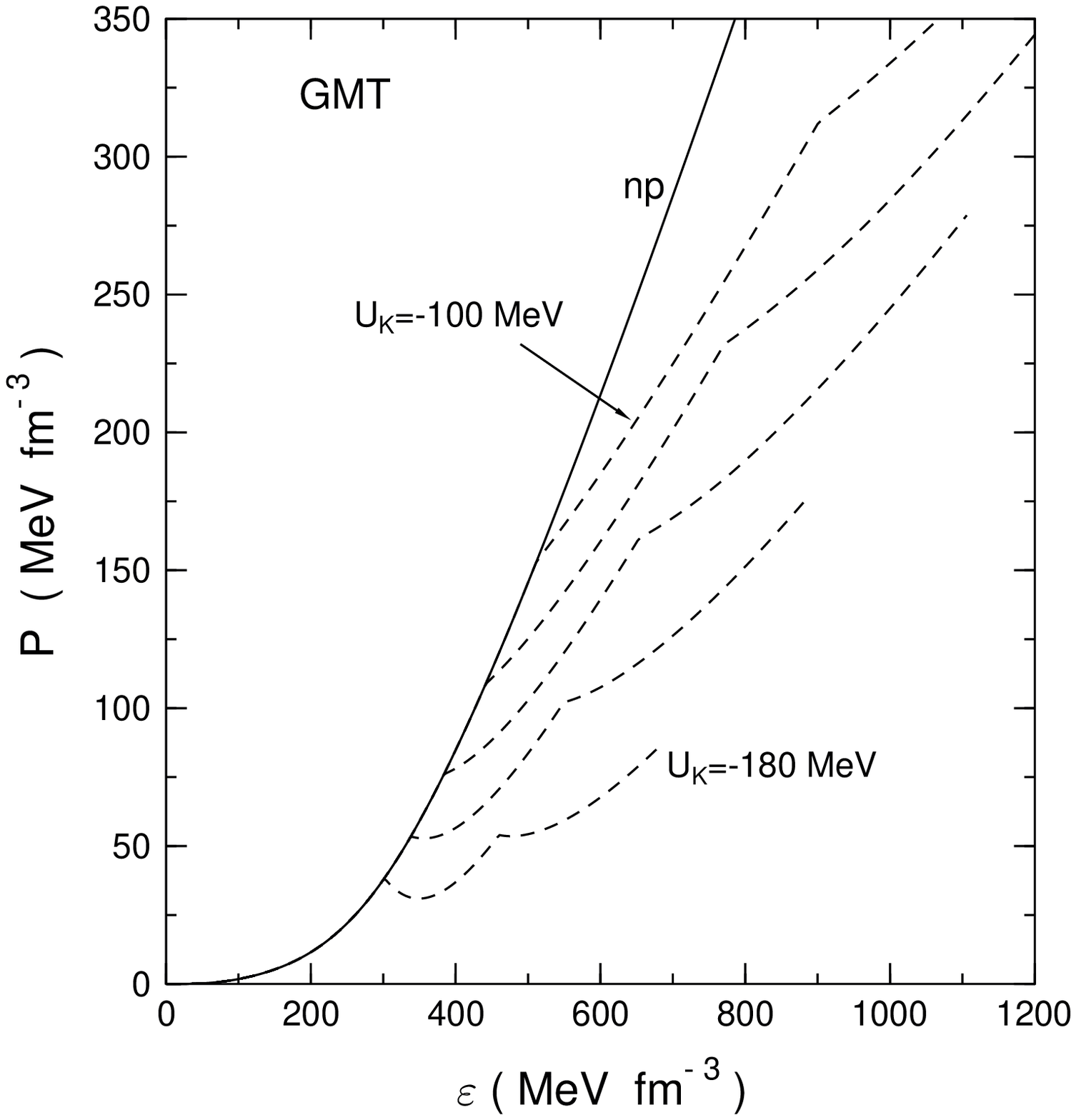}
}}

\vspace{-5cm}

\noindent{\small{
FIG. 10. Same as Fig. 5 but for the GMT set. The values of $U_{\bar K}(n_0)$
are same as in Fig. 9.}}
 \newpage 
\vspace{-2cm}

{\centerline{
\epsfxsize=14cm
\epsfysize=17cm
\epsffile{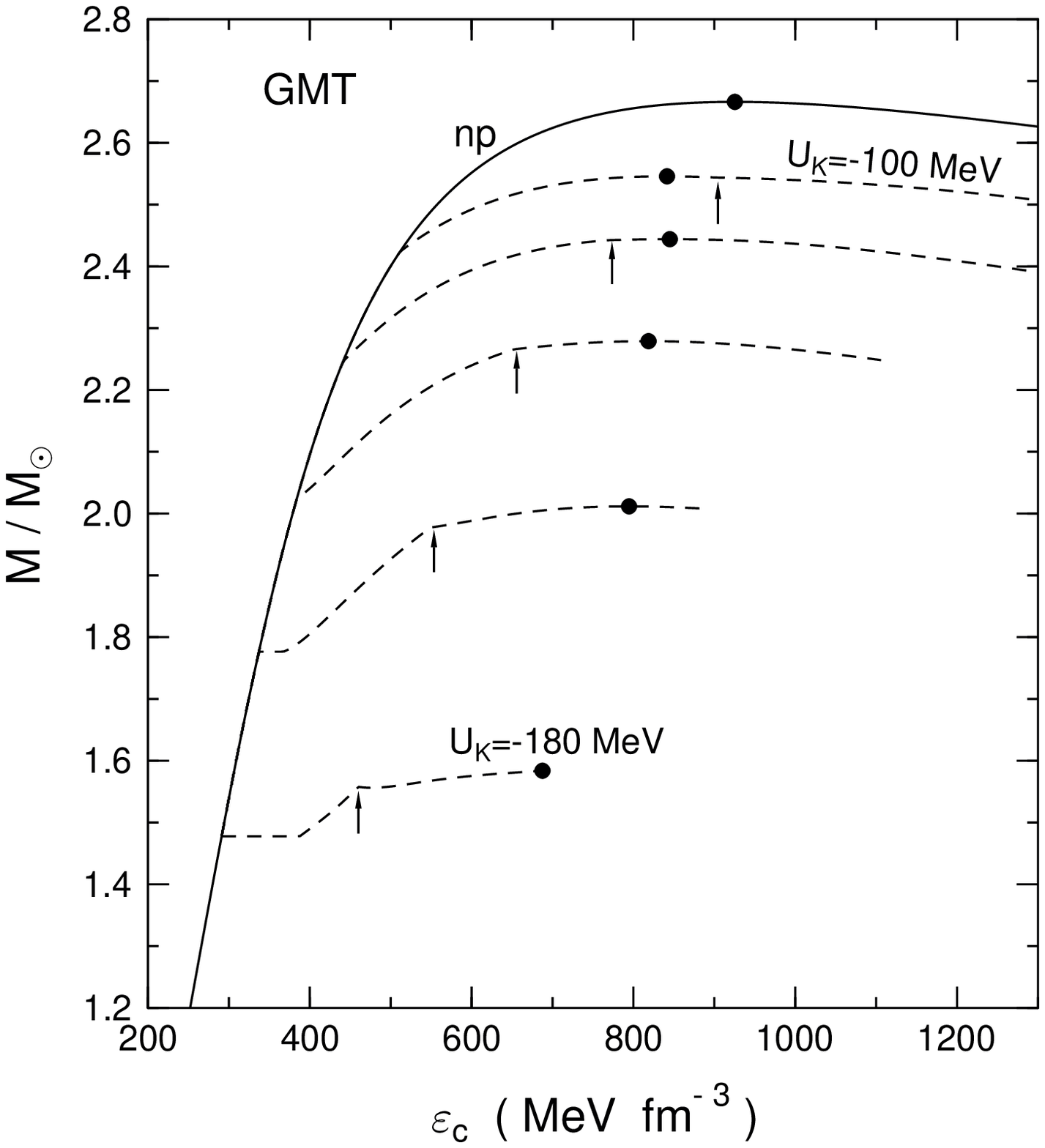}
}}

\vspace{-5cm}

\noindent{\small{
FIG. 11. Same as Fig. 6 but for the GMT set. The values of $U_{\bar K}(n_0)$
are same as in Fig. 9.}}
 \newpage 
\vspace{-2cm}

{\centerline{
\epsfxsize=14cm
\epsfysize=22cm
\epsffile{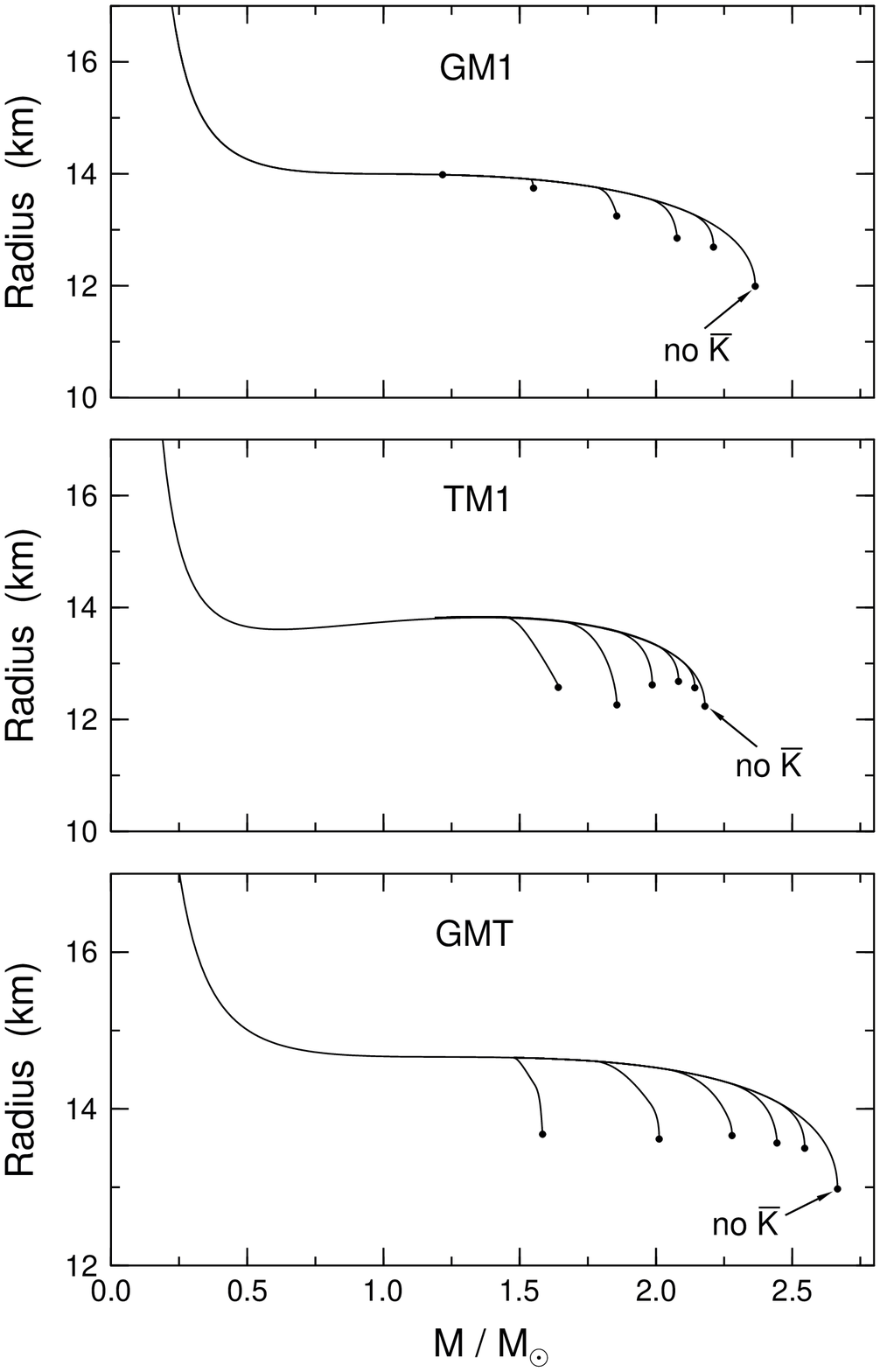}
}}

\vspace{-2.cm}

\noindent{\small{
FIG. 12. The mass-radius relation for neutron star sequences for 
nucleons-only stars and for stars with further inclusion of $K^-$ and 
$\bar K^0$ condensation for the GM1 (top panel), TM1 (central panel), and 
GMT (bottom panel) parameter sets. The filled circles correspond to the 
maximum masses of the stars without and with $\bar K$ condensate. For the 
latter case, the circles from right to left in each panel are obtained with 
$U_{\bar K}(n_0)= -100, -120, -140, -160, -180$ MeV.}}
 \newpage 
\vspace{-2cm}

{\centerline{
\epsfxsize=14cm
\epsfysize=22cm
\epsffile{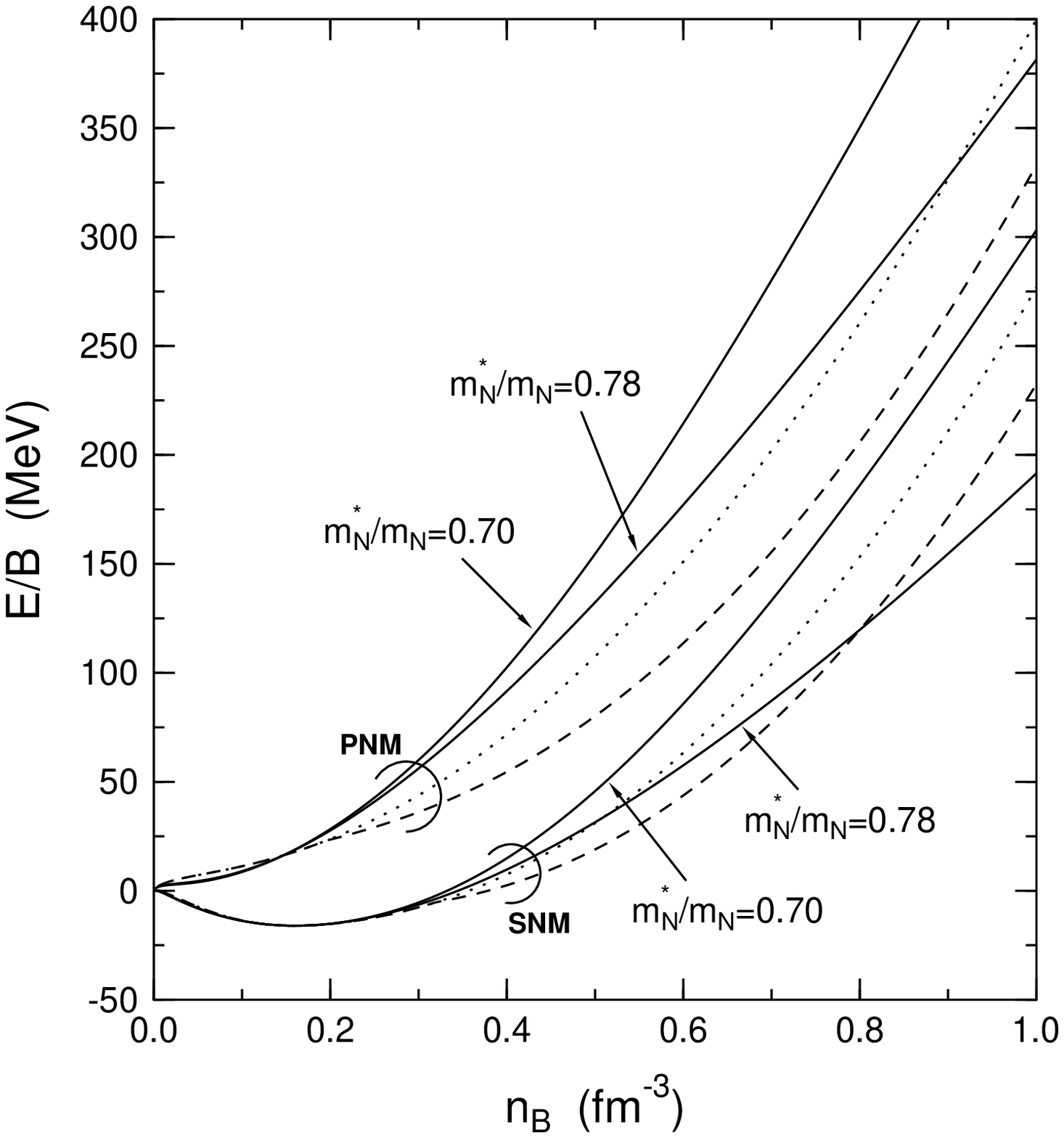}
}}

\vspace{-5cm}

\noindent{\small{
FIG. 13. The pure neutron matter (PNM) and symmetric nuclear matter (SNM) 
energies as a function of baryon density. The calculations are in the RMF 
model (solid lines) with effective nucleon mass of $m^*_N/m_N = 0.70$ and 0.78, 
and in the APR model [18] with pion condensation (dashed lines) and their
low density extrapolations without a pion condensed phase (dotted lines); 
for details see text.}}

\end{document}